\documentclass[12pt]{article}
\usepackage{epsfig}
\usepackage{graphics}
\usepackage{psfrag}
\usepackage{amsmath}
\usepackage{amssymb}

\newcommand{\bea}{\begin{eqnarray}}
\newcommand{\eea}{\end{eqnarray}}
\newcommand{\simgt}{\hbox{ \raise3pt\hbox to 0pt{$>$}\raise-3pt\hbox{$\sim$} }}
\newcommand{\simlt}{\hbox{ \raise3pt\hbox to 0pt{$<$}\raise-3pt\hbox{$\sim$} }}

\newcommand{\clfn}{\setcounter{footnote}{0}}

\begin{document}

\begin{titlepage}

    \begin{flushright}
      \normalsize TU-926\\
      \today
    \end{flushright}

\vskip1.5cm
\begin{center}
\Large\bf\boldmath
Algorithms to Evaluate Multiple Sums\\
for Loop Computations
\unboldmath
\end{center}

\vspace*{0.8cm}
\begin{center}
\large
{C. Anzai} and
{Y. Sumino}\\[5mm]
  {\small\it Department of Physics, Tohoku University}\\[0.1cm]
  {\small\it Sendai, 980-8578 Japan}

\end{center}

\vspace*{0.8cm}
\begin{abstract}
\noindent
We present algorithms to evaluate two types of
multiple sums, which appear in higher-order
loop computations. 
We consider expansions of
a generalized hyper-geometric-type sums,
$
\sum_{n_1,\cdots,n_N}
\frac{
\Gamma(\boldsymbol{a}_1\cdot\boldsymbol{n}+c_1)
\Gamma(\boldsymbol{a}_2\cdot\boldsymbol{n}+c_2)
\cdots
\Gamma(\boldsymbol{a}_P\cdot\boldsymbol{n}+c_P)
}{
\Gamma(\boldsymbol{b}_1\cdot\boldsymbol{n}+d_1)
\Gamma(\boldsymbol{b}_2\cdot\boldsymbol{n}+d_2)
\cdots
\Gamma(\boldsymbol{b}_Q\cdot\boldsymbol{n}+d_Q)
}~
x_1^{n_1}\cdots x_N^{n_N}
\rule[-3mm]{0mm}{7.5mm}
~~
$ 
with 
$\boldsymbol{a}_i\! \cdot \!\boldsymbol{n}=
\sum_{j=1}^N a_{ij}n_j
\rule[-3mm]{0mm}{5mm}
$, etc., 
in a small parameter
$\epsilon$
around rational values
of $c_i$,$d_i$'s. 
Type I sum corresponds to the case where, 
in the limit $\epsilon\to 0$, the
summand reduces to a
rational function of $n_j$'s
times $x_1^{n_1}\cdots x_N^{n_N}$; 
$c_i$,$d_i$'s can depend on an external integer index.
Type II sum is a double sum ($N=2$), where 
$c_i,d_i$'s are half-integers or integers as $\epsilon\to0$
and $x_i=1$; 
we consider some specific cases where
at most six $\Gamma$ functions remain in the limit $\epsilon\to0$.
The algorithms enable evaluations of 
arbitrary expansion coefficients in $\epsilon$
in terms of $Z$-sums and
multiple polylogarithms
(generalized multiple zeta values). 
We also present applications of these algorithms.
In particular, Type I sums can be used to generate a
new class of relations among generalized multiple zeta values.
We provide a Mathematica package, in which  these
algorithms are implemented.

\vspace*{0.8cm}
\noindent

\end{abstract}


\vfil
\end{titlepage}

\newpage

\section{Introduction}
\label{s1}

During the last decades, there have been remarkable
developments in technologies for computing
higher-order radiative corrections in quantum field theories.
Applications of these computational
technologies extend from phenomenology
of particle physics to various fields of theoretical research;
see, for instance, reviews 
\cite{Kreimer:2000ja,SmirnovBook,
Bogner:2009uz,Bern:2009kf,Ellis:2011cr,Blumlein:2012gu,Bern:2012zz}.

It has become a standard procedure to reduce numerous
integrals, which appear in higher-order corrections, to
a small set of integrals (master integrals) using identities 
derived by integration by parts \cite{Chetyrkin:1981qh}
or other types of identities.
One of the themes of today's computational technologies
for evaluating master integrals concerns how to reduce
a class of transcendental functions of the form\footnote{
For instance, this type of sums appear in the method
of Mellin-Barnes integral representations
for Feynman parameter integrals, after
closing integral contours \cite{SmirnovBook}.
}
\bea
&&
F(c_1,\dots,c_P;d_1,\dots,d_Q;x_1,\dots,x_N)
\nonumber\\&&
~~~
=
\sum_{n_1,\cdots,n_N}
\frac{
\Gamma(\boldsymbol{a}_1\! \cdot \!\boldsymbol{n}+c_1)
\Gamma(\boldsymbol{a}_2\! \cdot \!\boldsymbol{n}+c_2)
\cdots
\Gamma(\boldsymbol{a}_P\! \cdot \!\boldsymbol{n}+c_P)
}{
\Gamma(\boldsymbol{b}_1\! \cdot \!\boldsymbol{n}+d_1)
\Gamma(\boldsymbol{b}_2\! \cdot \!\boldsymbol{n}+d_2)
\cdots
\Gamma(\boldsymbol{b}_Q\! \cdot \!\boldsymbol{n}+d_Q)
}~
x_1^{n_1}\cdots x_N^{n_N}
\label{TranscendentalFn}
\eea
with
$\boldsymbol{a}_i\! \cdot \!\boldsymbol{n}=a_{i,1}n_1+\cdots+a_{i,N}n_N$,
etc.
Generally, $c_i$'s and $d_i$'s depend linearly on a
small expansion parameter $\epsilon$, and the
expansion coefficients in $\epsilon$ need to be computed.
The goal of the reduction is an expression in terms of
some simpler mathematical objects whose properties
are well known and which are
amenable to  fast high-precision numerical evaluations.
Mathematical functions that have been
explored for this purpose are the
harmonic sum, nested harmonic sum \cite{Vermaseren:1998uu}, 
polylogarithm, 
harmonic polylogarithms (HPLs) \cite{Remiddi:1999ew},
 multiple polylogarithm  \cite{Goncharov},
and their generalizations \cite{Moch:2001zr,Ablinger:2011te};
special values of these functions, such as multiple
zeta values (MZVs) \cite{Broadhurst:1996az}
and their generalizations 
\cite{Broadhurst:1998rz,Moch:2001zr,Ablinger:2011te}, also 
take part.

The problem becomes more and more intricate as the
order of loops is raised or as the number of scales
involved increases.
Typically in eq.~(\ref{TranscendentalFn}), 
the summation multiplicity $N$ becomes large,
more $\Gamma$ functions appear,
more entangled combinations of the indices 
$n_1,\dots,n_N$
appear in
the arguments of $\Gamma$ functions, and
expansions of $c_i$'s and $d_i$'s around  
more complicated rational numbers and to higher orders
in $\epsilon$ become requisite.

There have been systematic studies on reduction
of single sums ($N=1$) of the form of eq.~(\ref{TranscendentalFn})
\cite{Davydychev:1999mq,
Moch:2001zr,Weinzierl:2004bn,Kalmykov:2004kg,Huber:2005yg,
Kalmykov:2006hu}.
By contrast, studies on reduction of
multiple sums ($N\geq 2$) are still premature:
The methods of recursion relations \cite{Bytev:2011ks},
of differential equations \cite{Kalmykov:2012rr},
and of summation \cite{Moch:2001zr,Blumlein:2010zv}
have been examined and applied to computations of
multi-loop integrals.
What class of multiple sums can be reduced?
And to which extent can they be reduced?
Which method is more efficient or general?
--- Studies on these questions seem to have a long way ahead.

In this paper, we present algorithms to evaluate
two types of multiple sums, 
each of which corresponds to an arbitrary expansion coefficient
in $\epsilon$ of
a sub-class
of the above multiple sum.
Type I sum corresponds to the case where the ratio of
the $\Gamma$ functions in the summand 
of eq.~(\ref{TranscendentalFn}) reduces
to a rational function of $n_j$'s in the limit
$\epsilon\to0$, with integer coefficients of $n_j$'s
in each linear polynomial factor.\footnote{
This means that: $P=Q$;
with an appropriate ordering of $\Gamma$ functions,
$(c_i-d_i)_{\epsilon\to0}$ are integers 
and $\boldsymbol{a}_i=\boldsymbol{b}_i$ for all $i$'s; 
$a_{i,j},b_{i,j}$ as well as $c_i,d_i|_{\epsilon\to0}$
are rational numbers.
}
Furthermore, $c_i,d_i$'s can depend on an external 
integer index $\nu$ linearly, through which the sum
becomes dependent on $\nu$.
The algorithm can reduce the expansion coefficients
at any order in $\epsilon$ and for any 
summation multiplicity.
The result of reduction is expressed in terms of
$Z$-sums (generalized nested harmonic sums)
and their values at infinity 
(multiple polylogarithms), which depend on $x_1,\dots,x_N$.

Type II sum is a double sum with the form
$
\sum_{\ell,m=0}^{\infty}
\prod_{p}
\frac{\Gamma\left(\ell+a_p\right)}{\Gamma\left(\ell+b_p\right)}
\prod_{q}
\frac{\Gamma\left(m+c_q\right)}{\Gamma\left(m+d_q\right)}
\prod_{r}
\frac{\Gamma\left(\ell+m+e_r \right)}{\Gamma\left(\ell+m+f_r\right)}
$,
where $a_p,\dots,f_r$ are half-integers (including integers), and
for $p,q,r\geq 2$, the ratios of $\Gamma$ functions
reduce to rational functions of $\ell,m$ as $\epsilon\to0$.
This means that at most six $\Gamma$ functions remain as $\epsilon\to0$.
The algorithm can reduce the expansion coefficients
at any order in $\epsilon$.
The result of reduction is expressed in terms of
generalized MZVs.
We give precise definitions for the two types of sums 
in the main body of the paper.

Each type of sum covers a broad class of multiple sums
and has useful applications.
In particular, we can use Type I sum in the following ways:
(a) to convert a non-nested integral to a combination of
nested integrals, and
(b) to derive a new type of relations among the generalized
MZVs, which can be useful in reducing
expressions of final results.
We also apply the algorithms to compute  
master integrals for a static QCD potential at
three-loop order.
We present two new results.

Our algorithm for evaluating Type I sum 
is similar to the one developed in \cite{Blumlein:2010zv}
(see also \cite{Schneider02solvingparameterized}),
which is an algorithm adapted to evaluating 
a wide class of multiple sums for loop computations. 
The algorithm of \cite{Blumlein:2010zv} uses
solutions to difference equations
for transforming a multiple sum to a combination of
nested sums (which may later be expressed by harmonic sums,
multiple polylogarithms, etc.)
The method works if there are solutions
to difference equations in the solution space in which
they are searched for.
At present, however, in many cases one has to 
make explicit trials to see
which class of sums can eventually be 
transformed to nested sums.
Specifically, it has been unknown whether Type I
sum can be converted to nested sums in general cases.
On the other hand,
in our algorithm, we convert Type I sum to 
nested sums using the method of differences combined with
the invariance of the summand under shifts of indices,
in a recursive manner.
This corresponds to constructing explicitly the solution
to the difference equations with appropriate initial conditions,
in terms of $Z$-sums and multiple polylogarithms,
in the general case.
Thus, our algorithm provides an existence proof of the
solutions to the difference equations
for Type I sums in general.
Furthermore, our algorithm evaluates the sums
fairly efficiently.

Type II sums can be regarded as 
generalizations of the sums considered 
in \cite{Kalmykov:2000qe,
Davydychev:2003mv,Weinzierl:2004bn,Kalmykov:2006hu},
for the special values corresponding
to $x_j=1$. 
These previous studies 
evaluate expansion coefficients of
similar transcendental sums around half-integer values
of their arguments (not necessarily for $x_j=1$), 
but they
consider the cases where at most one pair of
$\Gamma$ functions remains
in the summand after expansion
(binomial sums or inverse-binomial sums).
In contrast,
we consider the cases where at most three pairs of $\Gamma$ functions remain,
whose arguments include entanglement of 
the indices essential to a double sum.
This adds significant complexity for evaluation of the sums
in comparison to the previous works.

In higher-order loop computations, more and more complicated classes of transcendental
functions (and their special values) appear.
In view of this general tendency, we use 
somewhat loose terminology in referring to
generalizations of MZVs and HPLs, 
such as colored MZVs, cyclotomic numbers, etc.
We simply refer to them as generalized MZVs or simply MZVs
hereafter, and in the case that we 
focus particularly on their functional dependence on $\beta_1$
(the upper bound of the outermost integral in integral representation),
we refer to them as (generalized) HPLs, although they are essentially 
the same quantities; they also belong to
a certain class of multiple polylogarithms.
See App.~A for the definitions.

The paper is organized as follows.
In Sec.~2, we present the algorithm for reduction of
Type I sums (Algorithm I).
Some applications of this algorithm are shown in Sec.~3.
We present the algorithm for reduction of
Type II sums (Algorithm II) in Sec.~4.
Sec.~5 provides applications of Algorithm II.
Summary and discussion are given in Sec.~6.
We deliver definitions and conventions in App.~A.
Some details of Algorithm II are explained in App.~B.

\section{\boldmath
Algorithm I: Multiple Sum without $\Gamma$ Functions}
\label{s2}
\clfn

In this section we present an algorithm (Algorithm I) to reduce
an $N$-ple sum of the following form:\footnote{
Note that polygamma functions, originating from expansions
of $\Gamma$ functions in eq.~(\ref{TranscendentalFn}), 
can be expressed using infinite series of rational functions.
}
\bea
S(\nu) = 
\sum_{i_1=a_1(\nu)}^{b_1(\nu)}
\sum_{i_2=a_2(\nu,i_1)}^{b_2(\nu,i_1)}
\cdots \sum_{i_N=a_N(\nu,i_1,\dots,i_{N-1})}^
{b_N(\nu,i_1,\dots,i_{N-1})}
~
\frac{\lambda_1^{i_1}\cdots \lambda_N^{i_N}}
{\prod_r L_r(\nu,i_1,\dots ,i_N)^{p_r}} ,
\label{SumAlg1}
\eea
where $\nu$ is an integer.
This includes, as a special case, in which
$S$ does not depend on $\nu$.
In eq.~(\ref{SumAlg1}),
$\lambda_n\in\mathbb{C}$;
each factor in the denominator $L_r(\nu,i_1,\dots ,i_N)$
is a linear polynomial of $\nu$ and 
$i_1,\dots ,i_N$ with integer coefficients\footnote{
Throughout this paper, ``a linear polynomial of $x_1,\dots,x_n$ with 
integer coefficients'' represents\\
$c_0 + c_1x_1 + \cdots + c_n x_n$ with $c_i \in \mathbb{Z}$.
}; $p_r\in \mathbb{N}$;
the upper or lower bound in each summation, $a_n$ or $b_n$, is
either infinity or a linear polynomial of its arguments
with integer coefficients.
We assume that, if $\nu$ is within an appropriate range,
the sum $S(\nu)$ is finite, namely, the summand is finite for any 
values of the indices and the sum converges if
some of $a_n,b_n$ are infinity.
After the reduction $S(\nu)$ is expressed in
terms of simple nested sums 
[generalized MZVs and $Z$-sums].

For clarity, we give 
simple examples:\footnote{
The second example can be expressed also in terms of polylogarithm
${\rm Li}_n(x)=\sum_{k=1}^\infty\frac{x^k}{k^n}$ via
$$
{\rm Im}[ Z(\infty;2,1;-i,i) ]=
-{\rm Im}\Bigl[\text{Li}_3\Bigl({
\frac{1+i}{2}}\Bigr)\Bigr]+\frac{\pi ^3}{128}+\frac{1}{32} \pi  \log ^2 2
.
$$
}
\bea
&&
\sum_{k=1}^{\infty}\sum_{m=1}^{\infty}
\frac{1}{(k+m)^2(2m+4k+1)}=
4-2{\cal C}\pi - \frac{\pi^2}{6}-4\log 2 + \frac{21}{4}\zeta(3) ,
\\
&&
\sum_{k=1}^{\infty}\sum_{m=1}^{k}\frac{(-1)^{k+m}}
{(k+1)^2(2m+1)}=
-8 \,{\rm Im}[ Z(\infty;2,1;-i,i) ]+4{\cal C}\log2 
-\frac{\pi ^3}{8}-\frac{\pi ^2}{12},
\\ &&
\sum_{n=0}^\infty \sum_{m=0}^\infty \frac{1}
{(1+n+\nu)^2 (1+m+n+\nu) (m+n+1)} 
\nonumber \\
&&
~~~~~~~~~~~~~~~~~~~
=
\frac{2}{\nu} Z(\nu;3;1) + \frac{1}{\nu}Z(\nu;1,2;1,1)-\frac{\pi^2}{6\nu}
Z(\nu;1;1) ,
\eea
where 
${\cal C}=\sum_{k=0}^\infty \frac{(-1)^k}{(2k+1)^2}$
denotes the Catalan constant;
$\zeta(z)=\sum_{n=1}^\infty \frac{1}{n^z}$ 
denotes the Riemann zeta function;
$Z(\infty;a_1,\cdots,a_N;b_1,\cdots,b_N)$ 
and $Z(\nu;a_1,\cdots,a_N;b_1,\cdots,b_N)$ represent
a (generalized) MZV and $Z$-sum,
respectively
(see App.~\ref{appA} for definitions).

\global\long\def\sumprime{\sideset{}{'}\sum}
It is convenient to use the generalized sum defined by\footnote{%
The generalization is convenient for relating sums to
functions such as
harmonic sums and polygamma functions.
For example,
${\sum'}_{k=0}^{n}\, \frac{1}{k+{1}/{2}}
=\mathrm{H}_{\frac{1}{2}+n}+\log 4$ holds for
an arbitrary integer $n$.%
}
\[
\sumprime_{k=a}^{b}f\left(k\right)=\begin{cases}
{\displaystyle ~~~\sum_{k=a}^{b}f\left(k\right)} & \left(a\leq b\right)
\rule[-7mm]{0mm}{6mm}
\\
~~~~~0 & \left(a=b+1\right)\rule[-3mm]{0mm}{6mm}
\\
{\displaystyle -\sum_{k=b+1}^{a-1}f\left(k\right)} & \left(a\geq b+2\right)
\end{cases}
\]
for integers $a$ and $b$.
The relations
\begin{gather*}
\sumprime_{k=b}^{a}f\left(k\right)=-\sumprime_{k=a+1}^{b-1}f\left(k\right)
,
\\
\sumprime_{k=a+\alpha}^{k=b+\beta}f\left(k\right)
=\sumprime_{k=a}^{b}f\left(k\right)+\sumprime_{k=b+1}^{b+\beta}f\left(k\right)
-\sumprime_{k=a}^{a+\alpha-1}
f\left(k\right)
\end{gather*}
hold for arbitrary integers
$a,b,\alpha$ and $\beta$.
Throughout this paper, it is understood that $\sum$ stands for
$\sum'$.

Let us define an integer matrix $A$ from the coefficients of
$L_r(\nu,i_1,\dots ,i_N)$'s by
\bea
A_{rn}=\frac{\partial L_r}{\partial i_n} .
\eea
Without loss of generality, we may assume that the rank of 
$A$ equals the number of the summation indices $N$.
In fact,
if ${\rm rank}(A)<N$, there exist redundant summation
indices, 
namely, with an appropriate redefinition of summation indices,
the summand can be rendered independent of 
$N\!-\!{\rm rank}(A)$ indices, and the summation over these indices
can be taken trivially.

Algorithm I consists of the following steps:
\begin{enumerate}
\item
Introduce regularization parameters if necessary and
perform partial fraction decomposition of the summand.
\item
Convert a multiple sum to a combination of nested sums.
This is achieved by
the method of differences for taking a sum of series,
combined with appropriate shifts of the summation indices.
\item
Remove regularization parameters and convert the 
nested sums to MZVs and 
$Z$-sums.
\item
Reduce MZVs and $Z$-sums to simpler ones.
\end{enumerate}
We explain each step.
\\

\noindent
\subsubsection*{Step 1}

We apply partial fraction decomposition to the summand of $S(\nu)$
with 
respect to each summation index $i_n$, starting from the innermost sum, 
up to the outermost sum.
By this procedure, $S(\nu)$ is rendered to a 
combination of the sums of the form eq.~(\ref{SumAlg1})
(apart from coefficients independent of $i_1,\dots,i_N$), 
with $N$ factors in the denominator,
$L_1, \dots, L_N$.
Here,
$L_1$ is independent of $i_2,\dots,i_{N}$,~
$L_{2}$ is independent of $i_3,\dots,i_{N}$,~ 
$\dots$,~ and
$L_{N-1}$ is independent of $i_N$.
The matrix $A$ is therefore rendered to an $N\times N$ lower
triangular matrix.

The partial fraction decompositions may
generate divergent terms.
By way of example:
\bea
\sum_{i_1,i_2,i_3}\frac{1}
{(i_1+i_2)(i_1+i_3)(i_2+i_3)}=
\sum_{i_1,i_2,i_3}\frac{1}{2i_1}
\left[\frac{1}{i_1+i_2}+\frac{1}{i_1-i_2}\right]
\left[\frac{1}{i_2+i_3}-\frac{1}{i_1+i_3}\right].
\eea
The term $1/(i_1-i_2)$ is divergent when $i_1=i_2$,
even if we restrict $i_1,i_2,i_3$ to be positive.
In general cases, if divergent terms are generated, 
we introduce regularization
parameters $\delta_n$ and shift\footnote{
The advantage of this regularization is that the limits $i_1\to i_2$
and $\delta_1,\delta_2\to 0$ commute.
}
$i_n \to i_n+\delta_n$.
At a later stage of the computation (step 3), we expand 
the result of the computation in $\delta_n$.
All singular terms in $\delta_n$ will cancel out.\footnote{
For practical implementation, it is often
more efficient to introduce a
single regularization parameter $\delta$ and to shift
$i_n \to i_n+c_n\delta$ with an appropriate choice of 
constants $c_n$'s.
}

\subsubsection*{Step 2}

After step 1, we obtain a combination of sums, each of
which has a form
\bea
S_2(\nu;\boldsymbol{\delta})=
\sum_{i_1=a_1(\nu)}^{b_1(\nu)}
\sum_{i_2=a_2(\nu,i_1)}^{b_2(\nu,i_1)}
\cdots \sum_{i_N=a_N(\nu,i_1,\dots,i_{N-1})}^
{b_N(\nu,i_1,\dots,i_{N-1})}
\frac{\lambda_1^{i_1}\cdots \lambda_N^{i_N}}
{\prod_{r=1}^N L_r(\nu,i_1,\dots ,i_N;\boldsymbol{\delta})^{p_r}} ,
\label{ResStep1}
\eea
up to an overall coefficient independent of 
$i_1,\dots,i_N$.
Here, $\boldsymbol{\delta}=(\delta_1,\dots,\delta_N)$.
The goal of step 2 is to convert $S_2$ to a
combination of nested sums;
a nested sum has a general form
\bea
\sum_{\Lambda(\nu)\geq j_{1}>\cdots>j_{M}\geq1}
f_1(j_1,\boldsymbol{\delta})f_2(j_2,\boldsymbol{\delta})
\cdots
f_M(j_M,\boldsymbol{\delta}),
\label{NestedForm}
\eea
where $\Lambda(\nu)$ is either $\infty$ or a linear polynomial of $\nu$
with integer coefficients.
We construct an algorithm by induction.
Suppose that for multiplicity $N\leq K-1$
we have a procedure to
convert the sum given by eq.~(\ref{ResStep1}) to a combination of
nested sums.
Below we show how to convert the sum with multiplicity $N=K$.

We first find a set of integers
$(\Delta_0,\Delta_1,\dots,\Delta_N)$ which satisfy
the following two conditions:
\begin{enumerate}
\renewcommand{\labelenumi}{(\roman{enumi})}
\item
For all $L_r$, 
\bea
L_r(\nu+\Delta_0,i_1+\Delta_1,\dots,i_N+\Delta_N;\boldsymbol{\delta})=
L_r(\nu,i_1,\dots,i_N;\boldsymbol{\delta}) .
\eea
\item
$\Delta_0 \neq 0$.
\end{enumerate}
From the condition (i), it follows that
$\sum_{n}A_{rn}\Delta_n+C_r\Delta_0=0$, where
$C_r=\partial L_r/\partial \nu$.
Hence,
\bea
\Delta_n = -\sum_r (A^{-1})_{nr}\,C_r\Delta_0 .
\eea
Since $(A^{-1})_{nr}$ are rational numbers, 
with an appropriate integer $\Delta_0\neq 0$,
all $\Delta_n$ can be set to integers.
We denote
\bea
&&
\alpha_n=a_n(\nu+\Delta_0,i_1+\Delta_1,\dots,i_{n-1}+\Delta_{n-1})-
a_n(\nu,i_1,\dots,i_{n-1}) ,
\\&&
\beta_n=b_n(\nu+\Delta_0,i_1+\Delta_1,\dots,i_{n-1}+\Delta_{n-1})-
b_n(\nu,i_1,\dots,i_{n-1}) ,
\eea
which are also integers.
\clfn

Using the property (i), one can relate 
$S_2(\nu+\Delta_0)$ to $S_2(\nu)$.
In fact, if we define the ``difference'' of the
two terms as
\bea
\Delta S_2(\nu) \equiv S_2(\nu+\Delta_0)-
\lambda S_2(\nu) 
~~~~;~~~~
\lambda=\prod_n \lambda_n^{\Delta_n} ,
\eea
we may rewrite
\bea
  && \Delta S_2(\nu) =
\lambda
\left[\sum_{i_{1}=a_{1}+\alpha_{1}}^{b_{1}+\beta_{1}}\cdots\sum_{i_{N}=a_{N}+\alpha_{N}}^{b_{N}+\beta_{N}}-\sum_{i_{1}=a_{1}}^{b_{1}}\cdots\sum_{i_{N}=a_{N}}^{b_{N}}\right]
\frac{\lambda_1^{i_1}\cdots \lambda_N^{i_N}}
{\prod_{r} L_r(\nu,i_1,\dots ,i_N;\boldsymbol{\delta})^{p_r}} 
~~~
\label{DeltaS2-2}
\eea
by shifting the indices as $i_n \to i_n+\Delta_n$ in
$S_2(\nu+\Delta_0)$.
With decomposition of the sums\footnote{
The decomposition may generate divergent terms.
In this case, one needs to introduce regularization parameters
as in step 1 (if not introduced already).
} 
\bea
 && 
 \sum_{i_{1}=a_{1}+\alpha_{1}}^{b_{1}+\beta_{1}}\cdots\sum_{i_{N}=a_{N}+\alpha_{N}}^{b_{N}+\beta_{N}}-\sum_{i_{1}=a_{1}}^{b_{1}}\cdots\sum_{i_{N}=a_{N}}^{b_{N}}
\nonumber\\
 &&~~
=
\left(\sum_{i_{1}=a_{1}}^{b_{1}}-\sum_{i_{1}=a_{1}}^{a_{1}+\alpha_{1}-1}+\sum_{i_{1}=b_{1}+1}^{b_{1}+\beta_{1}}\right)\cdots\left(\sum_{i_{N}=a_{N}}^{b_{N}}-\sum_{i_{N}=a_{N}}^{a_{N}+\alpha_{N}-1}+\sum_{i_{N}=b_{N}+1}^{b_{N}+\beta_{N}}\right)-\sum_{i_{1}=a_{1}}^{b_{1}}\cdots\sum_{i_{N}=a_{N}}^{b_{N}} ,
\nonumber\\
 &&~~
 \label{BulkCancel}
\eea
bulk of the difference
cancels.
The remaining terms include at least one ``surface term,''
$\sum_{i_n=a_n}^{a_n+\alpha_n-1}$
or $\sum_{i_n=b_n+1}^{b_n+\beta_n}$, whose sum can be evaluated explicitly
since $\alpha_n,\beta_n$ are explicit numbers.
Thus, the remaining terms have summation multiplicity $K-1$ or less.
According to the assumption of induction,
$\Delta S_2(\nu)$ can be expressed by a combination of nested sums,
eq.~(\ref{NestedForm}).

We can reconstruct $S_2(\nu)$ by summing up an initial term
$S_2(\nu_0)$ and the differences $\Delta S_2(\nu)$ with
appropriate weights.
Here, $\nu_0$ is the remainder in division of $\nu$ 
by $\Delta_0$, i.e., $\nu \equiv \nu_0 ~({\rm mod}~ \Delta_0)$ with
$0\leq \nu_0 < \Delta_0$.
We have
\bea
S_2(\nu)=\lambda^{(\nu-\nu_0)/\Delta_0}S_2(\nu_0) + 
\sum_{\mu=\nu_0}^{\nu-\Delta_0}
{\rm Proj}(\mu,\nu_0,\Delta_0)\,\Delta S_2(\mu) \,
\lambda^{(\nu-\mu-\Delta_0)/\Delta_0} .
\label{S2resum}
\eea
The projector is defined as
\bea
{\rm Proj}(m,n,d)  =\frac{1}{d}\sum_{k=1}^{d}
\exp\left(2\pi i k\frac{m-n}{d}\right)
  =\left\{
\begin{array}{ll}
0 & \text{if $m \not\equiv n~({\rm mod}~d)$}\\
1 & \text{if $m \equiv n~({\rm mod}~d)$}
\end{array}
\right. ,
\label{Proj}
\eea
such that the sum in eq.~(\ref{S2resum}) is taken
in steps $\Delta_0$.
Due to the presence of the projector, the upper and
lower limits of the summation in the second term of eq.~(\ref{S2resum})
can be replaced by $\nu-1$ and $0$, respectively.

In the first term of eq.~(\ref{S2resum}), if we 
assign a specific value to $\nu_0\in\{0,1,2,\dots,\Delta_0-1\}$,
$S_2(\nu_0)$ can be converted to a nested sum.
In fact, denoting $S_2(\nu_0)=\sum_{i_1}F(i_1)$,
we may regard $i_1$ as an external index
of a sum $F(i_1)$ of multiplicity $K-1$;
hence, $F(i_1)$ can be converted to a
combination of nested sums.

Since $\nu_0$ is a function of $\nu$ in 
eq.~(\ref{S2resum}),
for our purpose, we rewrite the right-hand side
in the following manner:
we replace $\nu_0$ by $\mu_0$, multiply by a
projector which projects out the term $\mu_0=\nu_0$,
and take a sum over $\mu_0$.
Hence,
\bea
&&
S_2(\nu)= \lambda^{\nu/\Delta_0}
\sum_{\mu_0=0}^{\Delta_0-1} {\rm Proj}(\nu,\mu_0,\Delta_0)
\nonumber\\&&
~~~~~~~~~~~~~
\times
\Biggl[
\lambda^{-\mu_0/\Delta_0}S_2(\mu_0) 
+ 
\sum_{\mu=0}^{\nu-1}
{\rm Proj}(\mu,\mu_0,\Delta_0)\,\Delta S_2(\mu) \,
\lambda^{-(\mu+\Delta_0)/\Delta_0} 
\Biggr].
\label{S2resum-2}
\eea
$\Delta_0$ being an explicit number, the sum over
$\mu_0$ can be evaluated explicitly, and the result is
$\lambda^{\nu/\Delta_0}$ times a combination of nested sums.
Thus, apart from $\nu$ dependent coefficients which multiply individual
terms, $S_2(\nu)$ is converted to a combination of nested sums.\footnote{
If the upper bound of a nested sum in $\Delta S_2$
includes an integer multiple of
the outer index ($\mu$), one can use the projector to synchronize the indices
and render the sum to a nested form:
$$
\sum_{\mu=0}^{\nu -1}\sum_{j=0}^{m\mu-1} f(\mu,j,\nu)=
\sum_{k=0}^{m\nu -1}\sum_{j=0}^{k-1} {\rm Proj}(k,0,m)\, f(k/m,j,\nu)\, ,
$$
where we set $k=m\mu$ for a given $m\in \mathbb{N}$.
The same manipulation is applied also to the first term $S_2(\mu_0)$ if necessary.
}

Let us present a simple example:
\bea
f(m)=\sum_{k=1}^\infty \frac{(-1)^k}{(1+k+m)^2}.
\label{ExStep2}
\eea
This can be expressed as
\bea
&&
f(m)=[f(m)+f(m-1)]-[f(m-1)+f(m-2)]
\nonumber\\&&
~~~~~~~~~~
+\cdots +
(-1)^m[f(2)+f(1)]-(-1)^mf(1)
\nonumber\\ &&
~~~~~~~
=\sum_{j=2}^m (-1)^{m-j}[f(j)+f(j-1)]-(-1)^mf(1).
\eea
Bulk of the sum in
the ``difference'' of two adjacent terms gets canceled 
since the shifts $m\to m-1,~k \to k+1$ leave the denominator
in eq.~(\ref{ExStep2}) unchanged:
\bea
f(j)+f(j-1) =\left(\sum_{k=1}^\infty - \sum_{k=2}^\infty\right)
\frac{(-1)^k}{(k+j)^2}
=\frac{-1}{(j+1)^2} .
\eea
Thus,
\bea
f(m)=(-1)^m\sum_{j=2}^m \frac{(-1)^{1-j}}{(j+1)^2}-(-1)^m
\sum_{k=1}^\infty \frac{(-1)^k}{(k+2)^2} .
\eea
In the first term, apart from the coefficient $(-1)^m$, dependence
on the external index $m$ enters only through the upper bound
of the summation, while the second term, apart from $(-1)^m$, is free of
$m$ and is essentially an MZV.

The above relation can be used to convert a double sum without
an external index to a combination of nested sums:
\bea
&&
\sum_{m=1}^\infty \sum_{k=1}^\infty 
\frac{(-1)^k \,i^m}{(2+m)^2(1+k+m)^2}
=\sum_{m=1}^\infty \frac{i^mf(m)}{(2+m)^2}
\nonumber\\&&
~~~
=-\sum_{m=1}^\infty
\frac{(-i)^m}{(2+m)^2}\sum_{j=2}^m \frac{(-1)^{j}}{(j+1)^2}-
\sum_{m=1}^\infty
\frac{(-i)^m}{(2+m)^2}
\sum_{k=1}^\infty \frac{(-1)^k}{(k+2)^2} .
\eea
The first term in the last line
is essentially a nested sum, while the second
term is a product of single sums.

\subsubsection*{Step 3}

We convert the nested sums obtained in step 2 to MZVs and $Z$-sums.
Up to an overall coefficient independent of the summation indices, 
each nested sum has a
form
\bea
S_3(\nu;\boldsymbol{\delta})=
\sum_{\Lambda(\nu)\geq j_{1}>\cdots>j_{M}\geq 0}
\frac{\lambda_1^{j_1}}
{(j_1+q_1+\boldsymbol{c}_1\cdot\boldsymbol{\delta})^{p_1}}
\times\cdots\times
\frac{\lambda_M^{j_M}}
{(j_M+q_M+\boldsymbol{c}_M\cdot\boldsymbol{\delta})^{p_M}} ,
\label{S3}
\eea
where $\boldsymbol{c}_n=(c_{n,1},\dots,c_{n,N})$ and
$c_{n,r}, q_n \in \mathbb{Q}$.
We have normalized the coefficient of each summation index 
$j_n$ to unity in
each factor in the denominator.

First we convert all the offsets $q_n$'s to integers.
Let $\ell$ be the least common multiple (LCM) of the denominators
of $q_1,\dots,q_M$.
In eq.~(\ref{S3}), we may replace $j_n$ by $j_n/\ell$ and 
insert ${\rm Proj}(j_n,0,\ell)$ for each $n$, such that 
each sum is taken on multiples of $\ell$:
\bea
S_3(\nu;\boldsymbol{\delta})=
\sum_{\ell \Lambda(\nu)\geq j_{1}>\cdots>j_{M}\geq 0}~
\prod_{n=1}^M 
\frac{{\rm Proj}(j_n,0,\ell)\,\lambda_n^{j_n}\,\ell^{p_n}}
{(j_n+\ell q_n+\ell \boldsymbol{c}_n\cdot\boldsymbol{\delta})^{p_n}} 
.
\label{MultiplyLCM}
\eea
Expanding the projector eq.~(\ref{Proj}), $S_3$ is given
by a combination of nested sums with integer offsets $\ell q_n$.

Next we shift each index $j_n\to j_n-\ell q_n$ to eliminate 
the offset.
By this procedure, divergent terms as $\delta_n\to 0$
are explicitly taken outside of the summation.
Then we expand in $\delta_n$'s [up to ${\cal O}(\delta_n^0)$]
the terms taken outside of the summation as well
as the summands.
All the divergent terms (which include negative powers of $\delta_n$'s)
cancel out.
Finally, by repeated applications of
shifts of summation indices, shifts of
upper and lower bounds of sums, 
and partial fraction decompositions, 
we can convert all the sums to
$Z$-sums [with argument $n\nu$ ($n\in\mathbb{N}$)]
and MZVs.

In the case that some of $\Lambda(\nu)$'s
are $\infty$, divergent MZVs as $j_n\to \infty$
may appear in individual sums.
By regularizing the sums carefully for large
values of $j_n$'s, these divergences can be extracted from
individual terms and they cancel out.
For instance, if we adopt a cut-off regularization, 
$\Lambda(\nu)=\infty \to \Lambda(\nu)=\Lambda\gg 1$, 
at most logarithmic divergences
$\sim (\log \Lambda)^P$ appear.
A shift of index $j_n$ alters
neither divergent part or the finite part as $\Lambda\to \infty$.
A multiplication of the cut-off in eq.~(\ref{MultiplyLCM})
induces a change
$(\log \Lambda)^P \to (\log \Lambda + \log \ell)^P$,
hence, this effect must properly be taken into account.

We give an example to illustrate the step 3:
\bea
&&
\sum_{k=1}^\nu \frac{1}{k-1+\delta}\sum_{m=1}^k
\frac{1}{m+1/2} =
\sum_{k=1}^{2\nu} \frac{{\rm Proj}(k,0,2)}{k/2-1+\delta}\sum_{m=1}^k
\frac{{\rm Proj}(m,0,2)}{m/2+1/2} 
\nonumber
\\ &&
=
\sum_{k=1}^{2\nu} \frac{1+(-1)^k}{k-2+2\delta}\sum_{m=1}^k
\frac{1+(-1)^m}{m+1}
=
\frac{2}{3\delta} +
\sum_{k=3}^{2\nu} \frac{1+(-1)^k}{k-2+2\delta}\sum_{m=1}^k
\frac{1+(-1)^m}{m+1}
\nonumber\\&&
=
\frac{2}{3\delta} +
\frac{52}{9}-\frac{2}{3\nu}-\frac{4}{3(1+2\nu)}-\frac{Z(2\nu;1;1)-Z(2\nu;1;-1)}{\nu}
\nonumber\\&&
~~~~~~
-2Z(2\nu;1;1)+\frac{10}{3}Z(2\nu;1;-1)+Z(2\nu;1,1;1,1)-Z(2\nu;1,1;1,-1)
\nonumber\\&&
~~~~~~
+Z(2\nu;1,1;-1,1)-Z(2\nu;1,1;-1,-1) + {\cal O}(\delta)
.
\eea
In the last equality, a
straightforward but somewhat cumbersome computation is needed
to convert the sum to a combination of $Z$-sums.

\subsubsection*{Step 4}

In many cases, host of
MZVs and $Z$-sums are included in the result of the
step 3.
They can be expressed by small sets of basis
elements for MZVs and $Z$-sums 
through specific reduction procedures or 
by utilizing known database.
Reductions of MZVs and $Z$-sums using various relations among them,
such as shuffle relations,
have been studied extensively in
the literatures, and we do not discuss them here;
see \cite{Blumlein:2009cf,Ablinger:2011te}
and references therein.
We apply such reduction procedures 
to simplify the result.
(We can also derive non-trivial relations among MZVs
using the algorithm given in this section.
This will be discussed in Sec.~\ref{s3b}.)
\\

\subsubsection*{Comment on surface terms at infinity}

In the case that some of the upper bounds of 
the summation in eq.~(\ref{SumAlg1})
are infinity,
the partial fraction decompositions in step 1 may
generate logarithmically divergent sums as $j_n\to\infty$.
This causes some subtleties to contributions of the
``surface terms,'' which appear in step 2.
For convergent sums, the surface terms from $j_n\to\infty$
certainly vanish.
It is {\it a priori} not obvious, however, whether contributions
of the surface terms
vanish in the case that there are individually divergent sums.
By way of example, consider a convergent sum ($=1$):
\bea
\sum_{k=1}^\infty\sum_{n=1}^\infty\frac{1}{k(k+n)(k+n+1)}
=
\sum_{k=1}^\infty\sum_{n=1}^\infty\frac{1}{k}
\left(\frac{1}{k+n}-\frac{1}{k+n+1}\right) .
\label{ExNonZeroST}
\eea
If we separate the sum on the right-hand
side, each sum becomes logarithmically divergent.
Let us replace the upper bounds by a cut-off $\Lambda\gg 1$
and convert the first sum $\sum_{k=1}^\Lambda \frac{f(k)}{k}$,
with $f(k)=\sum_{n=1}^\Lambda \frac{1}{k+n}$,
to a nested sum by
\bea
f(k)=f(1)+\sum_{j=2}^k [f(j)-f(j-1)]
=f(1)+\sum_{j=2}^k\left(\frac{1}{j+\Lambda}-\frac{1}{j}\right) .
\eea
Thus, the contribution of the surface term at $j=\Lambda$
is given by
\bea
\sum_{k=1}^\Lambda\frac{1}{k}\sum_{j=2}^k\frac{1}{j+\Lambda}
=\frac{\pi^2}{12} 
~~~~\mbox{as $\Lambda \to \infty$, }
\label{ExNonZeroST-2}
\eea
which is indeed non-vanishing.
It turns out that the contribution of the surface term from the second
term of eq.~(\ref{ExNonZeroST}) exactly cancels eq.~(\ref{ExNonZeroST-2}).
Hence, the contributions of the surface terms vanish as a whole.
This would not be surprising, since the sum eq.~(\ref{ExNonZeroST})
is convergent as $k\to\infty$.
Instead, if we introduce different cut-offs $\Lambda_k$ and $\Lambda_n$
for $k$ and $n$, respectively, and take the limit $\Lambda_n\to\infty$
first, we readily find that the individual surface terms vanish.

In general, this problem of the surface terms can be
treated properly by introducing cut-off
regularization as above.
In a particular regularization scheme and in the case
that all $|\lambda_n|=1$,\footnote{
Without this condition
the surface term in
eq.~(\ref{S2resum-2}), which contains $\lambda=\prod_n \lambda_n^{\Delta_n}$,
may be exponentially divergent at infinity, even for an
originally convergent sum with all $|\lambda_n|<1$.
(Note that $\Delta_n$ can be negative.)
}
we can argue that the surface terms from $j_n=\infty$ vanish 
(provided the computation is carried out in a specific
order, see below), so that these surface terms 
can be ignored in the computation.
This reduces the amount of computation considerably,
as compared to other regularization schemes, in which one has
to trace contributions of the surface terms through the computation.

The argument goes as follows.
Suppose that, 
in the original sum $S$ [eq.~(\ref{SumAlg1})],
the upper bounds of the summation indices
$i_{n_r}$ for $r=1,2,\dots,N'(\leq N)$ are infinity.
We replace the upper bounds of these indices by
$\Lambda_{n_r}$, respectively.
$S$ is convergent
as each $\Lambda_{n_r}\to\infty$.
For convenience, let us denote the external index
$\nu$ as $i_0$.
The surface terms are contained in $\Delta S_2$,
eq.~(\ref{DeltaS2-2}).
Since we apply the algorithm in step 2 recursively to
convert sums with lower multiplicities, the argument $i_x$ of
$S_2$ or
$\Delta S_2$ is not necessarily an external index ($i_0$)
but generally 
it can also be one of the summation indices $i_n$'s.
We can always work from the outermost to innermost sum
in applying the algorithm,
such that any of the indices corresponding to the surface terms
in $\Delta S_2$,
$i_{n}$, is {\it inner} with respect to the argument $i_x$ of
$\Delta S_2$, namely,
$x<n$ for all the summation indices $i_n$'s contained in $S_2(i_x)$.
Consider the contribution of only one surface term 
at $i_{n_r}=\Lambda_{n_r}$ to $\Delta S_2$:
it is suppressed by $\Lambda_{n_r}^{-p}$ with $p\geq 1$;
contributions of sums of indices other than
$i_{n_r}$ diverge at most logarithmically
$\sim \prod_{s\neq r} (\log \Lambda_{n_{s}})^{P_{s}}$.
If the index $i_x$ reaches up to order $\Lambda_x$,
the sum of $\Delta S_2(i_x)$ over $i_x$ is
at most order
$\Lambda_x/\Lambda_{n_r}^{p}\times
\prod_{s\neq r} (\log \Lambda_{n_{s}})^{P_{s}}$.
Therefore, if we parametrize all the cut-offs by a single parameter $u$
as $\Lambda_{n_r}=u^{n_r}$ and take a limit $u\to \infty$,
the cut-offs corresponding to inner indices diverge faster
($\Lambda_x\ll\Lambda_{n_r}$),
so that the contribution of the surface term vanishes.
(The outer sums modify the dependence on $u$ at most
logarithmically, which does not alter vanishing of
the surface term contribution.)
The contribution of more than one surface term vanishes
even faster with $u$.

To summarize, in the case
that all $|\lambda_n|=1$, if we work from the outermost sum to innermost
sum in converting non-nested
sums to combinations of nested sums
(if we do not exchange the order of summation indices),
we can neglect contributions of the surface terms at infinity.

\section{Applications of Algorithm I}
\label{s3}

\subsection{Conversion of Non-nested Integral to Nested Integrals}
\label{s3a}
\clfn 

The algorithm given in the previous section 
(Algorithm I) converts
a certain type of non-nested sums to combinations of
nested sums.
A straightforward application of this algorithm is to convert
a certain type of non-nested integrals to a combination
of nested integrals such as HPLs.
In fact, in problems suited to 
manipulations in integral representations,
it is often useful to convert a non-nested multiple
integral to a combination of nested integrals.

There are multiple integrals (with one external parameter $x$) 
which can be converted to multiple sums
of the form
\bea
F(x)=\sum_{\nu=1}^\infty \frac{x^\nu}{\nu^P}\, S(\nu)\,
{\rm Proj}(\nu,0,d)
~~~~;~~~~P\in\mathbb{N},
\eea
by first expanding the integrand in Taylor series and
then integrating.
Here, $S(\nu)$ is given by eq.~(\ref{SumAlg1}).
This may occur in the case that the integrand is a combination of
a rational function, logarithms, and HPLs of the integral variables.
We can use Algorithm I to convert $S(\nu)$
to a combination of nested sums;
in this form $F(x)$ is expressed as a combination of nested integrals
straightforwardly.
We give a simple example for illustration:
\bea
&&
I(x)=\int^x_0 dy\, \log(1-y/x)\,\log(1-xy)
=
\int^x_0 dy\,\sum_{k=1}^\infty\frac{(y/x)^k}{k}
\sum_{m=1}^\infty\frac{(xy)^m}{m}
\nonumber\\&&
~~~~~
= \sum_{m=1}^\infty \frac{x^{1+2m}}{m}
\sum_{k=1}^\infty \frac{1}{k(k+m+1)}
= \sum_{m=1}^\infty \frac{x^{1+2m}}{m}
\left[\frac{1}{(m+1)^2}+\frac{Z(m;1;1)}{m+1}
\right] .
\eea
In the last step we applied Algorithm I.
Using standard relations among MZVs and HPLs,
we may express the above result in terms of HPLs:
\bea
&&
I(x)=2x - \frac{1-x^2}{x}\Bigl[
Z(\infty;1;x^2)+Z(\infty;2;x^2)+Z(\infty;1,1;x^2,1)
\Bigr]
\nonumber\\&&
~~~
=2x + \frac{1-x^2}{x}\Bigl[
H_{-1}(x)+H_1(x)-H_{-1,-1}(x)-H_{-1,1}(x)
\nonumber\\&&
~~~~~~~~~~~~~~~~~~~~~~~~~
+2H_{0,-1}(x)+2H_{0,1}(x)-H_{1,-1}(x)-H_{1,1}(x)\Bigr] .
\eea
This is a combination of nested integrals;
see eqs.~(\ref{HPL1})--(\ref{HPL}) for the definition of HPL.

In more general cases,
one may need to redefine
the summation indices appropriately,
so that the exponent of $x$ is 
expressed in terms of a single index.
For instance, we redefine $k+n=l$ in
\bea
\sum_{k=1}^{\infty}\sum_{n=1}^{\infty}x^{k+n}F\left(k,n\right)
=
\sum_{l=2}^{\infty}x^{l}\,\sum_{n=1}^{l-1}F\left(l-n,n\right)
,
\eea
and convert the inner sum to a combination of $Z$-sums
with arguments $m l$ $(m\in\mathbb{N})$.

Examples of integrals, which can be converted to 
combinations of nested integrals in a similar manner,
and which may be useful, are given by

\bea
&&
I_1(x)=
\int^x_0 dy\, y^{p(1)}\, (\log y)^{p(2)} \, [\log(1-y)]^{p(3)}\, 
 [\log(1+y)]^{p(4)}  
\nonumber\\&&
~~~~~~~~~~~~~~~~~\times
[\log(1-y/x)]^{p(5)}\, 
 [\log(1-xy)]^{p(6)}\, H(y) ,
\label{ExI1}
\\ &&
I_2(w)=\int_0^w dx \int_0^x dy \int_0^y dz\,
x^{p(1)}y^{p(2)}z^{p(3)}
\left[\log \left(1 - {xy}/{w}\right)\right]^{p(4)}
\nonumber\\&&
~~~~~~~~~~~~~~~~~~~~~~~~~~~
\times
\left[\log \left(1 - {yz}/{w}\right)\right]^{p(5)}
\left[\log \left(1 - {zx}/{w}\right)\right]^{p(6)},
\eea
where $p(n)\in \mathbb{N}$, and $H(y)$ is an HPL.
The resulting expressions, however, may be quite lengthy.

\subsection{Relations among MZVs}
\label{s3b}

As already stated, a number of relations among
MZVs
have been derived and used for expressing MZVs by a
set of basis elements.
We show that Algorithm I can be used to derive
yet another type of relations among (generalized) MZVs.

We demonstrate it using an example.
Suppose that $\lambda\in\mathbb{C}$ satisfies
\bea
|\lambda|\leq 1~~~\mbox{and}~~~\lambda\neq 1.
\label{Condlambda}
\eea
We rearrange the order of the summation of a weight-two MZV:
\bea
&&
Z(\infty;1,1;\lambda,-\lambda)=\sum_{n>m>0}
\frac{\lambda^n(-\lambda)^m}{nm}
=\sum_{s=3}^\infty\sum_{m=1}^{\left\lfloor\frac{s-1}{2}
\right\rfloor}\lambda^s\,\frac{(-1)^m}
{m(s-m)}
\nonumber\\&&
~~~~~~
=
\sum_{r=1}^\infty \lambda^{2r+1} \sum_{m=1}^r
\frac{(-1)^m}{m(2r+1-m)}
+
\sum_{r=1}^\infty \lambda^{2r+2} \sum_{m=1}^r
\frac{(-1)^m}{m(2r+2-m)}
.
\eea
We set $s=n+m$. 
The upper bound of $m$ follows from the condition
$m\leq n-1=s-m-1$, where $\lfloor x \rfloor$
denotes the greatest integer less than or equal to $x$.
In the last line we separate $s$ to odd and even numbers
and set $s=2r+1$ and $s=2r+2$, respectively.
The sums over $m$ can be converted to combinations of
$Z$-sums by Algorithm I.
It can be shown that the above rearrangement of summation
is justified under the condition
eq.~(\ref{Condlambda}).
Thus, we find a relation among MZVs:
\bea
&&
z(\lambda,-\lambda)=
-\frac{1}{2}z(-2\lambda^2)+\frac{3}{4}z(-\lambda,-1)
-z(-\lambda,-i)-z(-\lambda,i)-\frac{1}{4}z(-\lambda,1)
\nonumber\\&&
~~~~~~~~~~~~~~
-\frac{1}{4}z(\lambda,-1)+z(\lambda,-i)+z(\lambda,i)
-\frac{1}{4}z(\lambda,1)+\frac{1}{4}z(\lambda^2,1).
\label{RelMZV}
\eea
We use a short-hand notation for MZV, eq.~(\ref{short-hand-not}),
where the original forms can be reproduced
via
$z(x_1,\dots,x_n)\to
Z(\infty;\alpha_1,\dots,\alpha_n;\beta_1,\dots,\beta_n)
$ with $\alpha_i = |x_i(\lambda=1)|$ and
$\beta_i=x_i/\alpha_i $.
The relation eq.~(\ref{RelMZV}) may be regarded as a special
case of the conversion relation described in Sec.~\ref{s3a}.

In particular, in the case that $\lambda^n \!=\!\pm i$
for some $n\in\mathbb{N}$, eq.~(\ref{RelMZV}) may lead to
an interesting relation.
We have examined independence of the above relation from other
relations, such as shuffle relations and those which can be
derived by variable transformations in integral representations
of MZVs
(e.g.\ $x\to x^{-1}$, $x\to x^2$,
$x\to \frac{1+ix^2}{1-ix^2}$).\footnote{
Using these relations 
we obtain, for instance, a relation between the 
basis elements at $w=3$ for the infinite cyclotomic harmonic sums
over
$\{(\pm1)^k/k,(\pm1)^k/(2k+1)\}$
proposed in \cite{Ablinger:2011te}:
$$
-\frac{1}{4}\sigma_{\{1,0,-2\},\{2,1,-1\}}+
\sigma_{\{2,1,-2\},\{1,0,-1\}}+
{\cal C} \log 2-\frac{\pi ^3}{32}+\frac{\pi
   ^2}{16}-1+\log 2 =0,
$$
hence, $\zeta(3)$ and 
${\rm Im}\bigl[{\rm Li}_3\bigl(\frac{1+i}{2}\bigr)\bigr]
\bigl(=
-{\cal C} \log 2-\frac{1}{2}\sigma_{\{2,1,-2\},\{1,0,-1\}}
+\frac{5 \pi ^3}{128}+\frac{\pi}{32}   \log ^2 2
\bigr)$
suffice to be introduced at this weight.
}
In the case that $\lambda=\pm i$, eq.~(\ref{RelMZV})
does not give a new relation.
On the other hand, in the case $\lambda = e^{i\pi/4}$
(a primitive 8th root of unity),
it gives a new relation, which can be used for reduction
of MZVs.
In fact, this gives a relation between the three basis elements 
at $w=2$, $l=8$ proposed in \cite{Ablinger:2011te}:
\bea
&&
-3 i\, {{\rm Im}[{\rm Li}_2(e_8)]}
+3 {\sigma_{1,1}(e_8,e_4)}
+{\sigma_{1,1}(e_8,e_8^3)}
+2 i {\cal C}+\frac{53 \pi ^2}{192}
\nonumber\\&&
~~
+\frac{1}{2}
   \log ^2\Bigl(\sqrt{2}\!-\!1\Bigr)-\frac{\log ^22}{8}-\frac{5}{4} \log 2 \,\log
   \Bigl(\sqrt{2}\!-\!1\Bigr)+\frac{9}{16} i \pi  \log 2
   =0
   .
\eea
Therefore, the number of the
(new) basis elements reduces to two.

It is easy to obtain
relations among MZVs in more complicated cases, by
rearranging summation orders using
this method.
The strategy is to rewrite an MZV 
$Z(\infty;\alpha_1,\dots,\alpha_n;\beta_1,\dots,\beta_n)$,
for which some of $|\beta_i|$'s are the same (equal to $|\lambda|$),
in a combination of HPLs of the type 
$H_{a_1,\dots,a_{n}}(\omega \lambda^m )$
(MZVs with $\beta_1=\omega \lambda^m $).
Here, $\omega$ denotes a root of unity.

Up to now, we do not know whether these relations 
can be derived from other known methods.
Extensive exploration of the
relations among MZVs for the case
beyond $\beta_i=\pm 1$ (e.g.\ $n$-th root of unity for $n>2$)
is still underway 
\cite{Broadhurst:1998rz,Borwein:2000et,Ablinger:2011te}.
We expect that the above method would be a useful tool for
analyses in this direction.
We have checked in specific examples that they 
lead to relations independent of shuffle relations
and are powerful in reducing MZVs to a small set of elements.
Generally, a huge set of relations need to be generated,
which requires an efficient algorithm for generation
of the relations, and Algorithm I is useful
in this respect.

\subsection{Evaluation of a 3-Loop Master Integral}
\label{s3c}

\begin{figure}[t]\centering
\includegraphics[width=5cm]{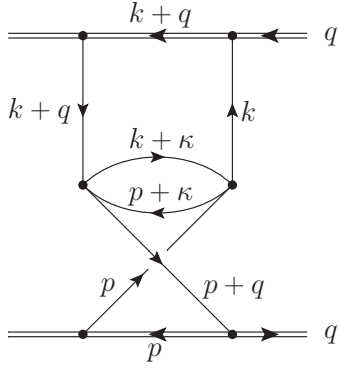}
\caption{\small
Diagram for the master integral ${\cal M}_1$.
Double lines represent heavy-quark propagators, while
single lines represent massless propagators.
\label{FigM33}
}
\end{figure}

We apply Algorithm I for evaluating a 3-loop
integral 
\bea
{\cal M}_1=\int \frac{d^Dp}{(2\pi)^D}
\frac{d^Dk}{(2\pi)^D}\frac{d^D\kappa}{(2\pi)^D}\,
\frac{1}{(k\cdot v)(p\cdot v)k^2p^2(k+q)^2(p+q)^2
(k+\kappa)^2(p+\kappa)^2} .
\label{M33}
\eea
This is one of the 40 master integrals, which appear in a computation
of the 3-loop correction to the static QCD potential
($\bar{a}_3$) \cite{Anzai:2009tm,Anzai:2010td};
the diagram is shown in Fig.~\ref{FigM33}.
In dimensional regularization ($D=4-2\epsilon$), expansion
coefficients of ${\cal M}_1$ in $\epsilon$ up to
(and including) order $\epsilon$ is necessary.
In eq.~(\ref{M33}), we omit $+i0$ in the propagator denominators,
hence, \{$k\cdot v$, $k^2$, $(k+q)^2$, $\dots$\} stand for
\{$k\cdot v + i0$, $k^2+i0$, $(k+q)^2+i0$, $\dots$\};
$q=(0,\vec{q})$ is a spacelike external momentum, whereas
$v=(1,\vec{0})$ is a temporal unit vector;
$(k\cdot v + i0)^{-1}$ represents a heavy-quark propagator. 

Using a standard technology \cite{SmirnovBook}, such as the one described
in \cite{Smirnov:2004ip}, one can convert the above integral to a 
three-fold Mellin-Barnes
integral.
Closing the contour in the complex plane for each Mellin-Barnes
integral
variable, one may further convert the integral to 
a combination of three-fold sums with $\Gamma$ functions.
By expanding in $\epsilon$,  at every order of $\epsilon$
all the $\Gamma$ functions
cancel, 
and we are left with sums of the form eq.~(\ref{SumAlg1}).
We evaluate the expansion coefficients up to ${\cal O}(\epsilon)$
using Algorithm I,
where sums with multiplicity up to six need to be evaluated.
The result is given by
\bea
&&
{\cal M}_1=A(\epsilon)(-q^2)^{-1-3\epsilon}\times
\left[
\frac{\pi ^4}{\epsilon } +
\biggl\{
-93 \zeta (5)-14 \pi
   ^2 \zeta (3)+\pi ^4 \Bigl( 2+6 \log 2 \Bigr)
\biggr\}
\right.
\nonumber\\&&
~~~~~~~~~~~~
+
\epsilon \, \biggl\{-288 s_6
-186 \zeta (5)+234 \zeta (3)^2
+\pi ^2 \Bigl(96 \text{Li}_4\bigl({\textstyle \frac{1}{2}}\bigr)
-28
   \zeta (3)+4 \log ^4 2\Bigr)
\nonumber\\&&
~~~~~~~~~~~~~~~~~~~
+\pi ^4 \Bigl(4+12 \log 2+14 \log ^2 2\Bigr)
+\frac{293 \pi   ^6}{420}
\biggr\}
\Biggr],
\\&&
A(\epsilon)=
-i \, (4\pi) ^{3 \epsilon -6} \exp(-3 \gamma_E  \epsilon ) ,
\eea
where $\gamma_E = 0.5772...$ denotes the Euler constant;
$s_6=z(-5,-1)+\zeta(6)=\sum_{m\geq n\geq 1}\frac{(-1)^{m+n}}{m^5n}=
0.9874414264...$;
${\rm Li}_n(x)=\sum_{k=1}^\infty\frac{x^k}{k^n}$ denotes 
polylogarithm.
We have checked the above result 
numerically by evaluating Feynman parameter integral
representation using sector decomposition.
This result has not been published elsewhere.

\section{\boldmath Algorithm II: Expansion Coefficients of
Double Sum with $\Gamma$ Functions}
\label{s4}
\clfn

We consider a double sum
\bea
&&
K(\boldsymbol{a},\boldsymbol{b};\boldsymbol{c},\boldsymbol{d};
\boldsymbol{e},\boldsymbol{f})
=
\sum_{\ell=0}^{\infty}\sum_{m=0}^{\infty}
\Biggl[\prod_{p=1}^L
\frac{\Gamma\left(\ell+a_p\right)}{\Gamma\left(\ell+b_p\right)}
\Biggr]
\Biggl[\prod_{q=1}^M
\frac{\Gamma\left(m+c_q\right)}{\Gamma\left(m+d_q\right)}
\Biggr]
\Biggl[\prod_{r=1}^N
\frac{\Gamma\left(\ell+m+e_r \right)}{\Gamma\left(\ell+m+f_r\right)}
\Biggr]
,
\nonumber\\&&
\label{eq:K}
\eea
where $\boldsymbol{a}=(a_1,\dots,a_L)$, etc.
In this section we present an algorithm (Algorithm II) for computing
expansion coefficients of
$K$ in a small parameter
around integer or half-integer arguments.
We assume the following conditions for
the arguments of $K$:
\begin{enumerate}
\renewcommand{\labelenumi}{(\roman{enumi})}
\item
$\boldsymbol{a}=\boldsymbol{a}_0+\epsilon\boldsymbol{a}_1$,
$\dots$,
$\boldsymbol{f}=\boldsymbol{f}_0+\epsilon\boldsymbol{f}_1$,
where $a_{0,p},b_{0,p},c_{0,q},d_{0,q},e_{0,r},f_{0,r}$ are 
half-integers (including integers) and
$\epsilon$ is a small expansion
parameter.
$a_{1,p},b_{1,p},c_{1,q},d_{1,q},e_{1,r},f_{1,r}$ 
are complex variables used to parametrize the expansion coefficients.
\item
For $p,q,r\geq 2$, $a_{0,p}-b_{0,p}$, $c_{0,q}-d_{0,q}$, 
$e_{0,r}-f_{0,r}$ are
integers.
\item
The sum in eq.~(\ref{eq:K}) is convergent at
$(\boldsymbol{a}_0,\boldsymbol{b}_0;\boldsymbol{c}_0,
\boldsymbol{d}_0;\boldsymbol{e}_0,\boldsymbol{f}_0)$, hence
$K$ is regular with respect to $\boldsymbol{a}_1,\dots,\boldsymbol{f}_1$
as $\epsilon\to 0$.
\end{enumerate}
The algorithm reduces
an arbitrary expansion coefficient of $K$ in 
$\epsilon$, up to any order, to a combination of MZVs.
Due to the above conditions, there remain at most six uncanceled
$\Gamma$ functions in the summand in
each expansion coefficient, before
reduction.

We give an example:
\bea
&&
\frac{\partial}{\partial\epsilon}\frac{\partial}{\partial\delta}\frac{\partial^{2}}{\partial\alpha^{2}}\,
C(\epsilon,\delta,\alpha)\!
\sum_{k,m=0}^{\infty}\frac{\Gamma(k\! + \!1)\Gamma(m\! + \!1)\Gamma(k\! + \!m\! + \!\frac{3}{2})\Gamma(k\! + \!m\! + \!\epsilon\! + \!2)}{\Gamma(k\! + \!\alpha\! + \!\frac{5}{2})\Gamma(m\! + \!\frac{5}{2})\Gamma(k\! + \!m\! + \!2)\Gamma(k\! + \!m\! + \!\delta\! + \!3)}
\Biggr|_{\epsilon,\delta,\alpha\to 0}
\nonumber\\
&&
=  -7296s_{6}-\frac{508\pi^{6}}{63}+\pi^{4}\left(-8\log^{2}2+\frac{32\log2}{45}+\frac{292}{15}\right)
+\pi^{2}\Biggl\{\zeta({3})\left(56\log2+\frac{656}{3}\right)
\nonumber\\
 && ~~~
+\frac{152\log^{4}2}{3}+192\log^{3}2+\frac{416\log^{2}2}{3}+\frac{32\log2}{3}+192\text{Li}_{4}\bigl({\textstyle \frac{1}{2}}\bigr)-\frac{80}{3}
\Biggr\}
+972\zeta({3})^{2}
\nonumber\\
&&~~~
 +\zeta({5})(6882\log2+22318)+\zeta({3})(-448\log^{3}2-2024\log^{2}2
-752\log2-248)
\nonumber\\
 && ~~~
-\frac{1024\log^{6}2}{15}-\frac{3808\log^{5}2}{15}-\frac{32\log^{4}2}{3}+\frac{1024\log^{3}2}{3}
+\log^{2}2\Bigl(-3072\text{Li}_{4}\bigl({\textstyle \frac{1}{2}}\bigr)+192\Bigr)
\nonumber\\
 && ~~~
 +\log2\Bigl(-9216\text{Li}_{5}\bigl({\textstyle \frac{1}{2}}\bigr)-8320\text{Li}_{4}\bigl({\textstyle \frac{1}{2}}\bigr)-160\Bigr)
\nonumber\\
&& ~~~
-12288\text{Li}_{6}\bigl({\textstyle \frac{1}{2}}\bigr)-11136\text{Li}_{5}\bigl({\textstyle \frac{1}{2}}\bigr)-256\text{Li}_{4}\bigl({\textstyle \frac{1}{2}}\bigr)-144
,
\eea
where we have included a prefactor 
$C(\epsilon,\delta,\alpha)
={\Gamma(\delta\! + \!3)\Gamma(\alpha\! + \!\frac{5}{2})}/
{\Gamma(\epsilon\! + \!2)}$
in order to eliminate trivial $\gamma_E$ and $\sqrt{\pi}$.

The algorithm consists of the following steps:
\begin{enumerate}
\item
Convert $K$ to an integral representation.
\item
Apply Kummer's formula and other variable transformations
to simplify the integral and to eliminate half-integer
powers in the integrand.
\item
Expand in $\epsilon$.
If necessary, regularize appropriately
before the expansion.
\item
Convert multiple integrals to combinations of
nested integrals.
Integrals with respect to individual
integral variables are converted recursively.
In each conversion process, we
extract singularities using integration by parts,
remove regularization parameters, and apply Algorithm I.
In the end the result is expressed by MZVs.
\item
Reduce MZVs to simpler ones.
\end{enumerate}
We explain each step below.
\\

\noindent
\subsubsection*{Step 1}

For later convenience, we denote 
\bea
e_r=e'_{N+1-r},~~~
f_r=f'_{N+1-r}
~~~~(1\leq r \leq N).
\label{efPrime}
\eea
We express each pair of $\Gamma$ functions using the
integral representation of the beta function as
\bea
&&
\frac{\Gamma\left(\ell+a_p\right)}{\Gamma\left(\ell+b_p\right)}
=\frac{1}{\Gamma(b_p-a_p)}\int_0^1 dx_p \, x_p^{\ell+a_p-1}
(1-x_p)^{b_p-a_p-1}
,
\label{Beta1}
\\&&
\frac{\Gamma\left(m+c_q\right)}{\Gamma\left(m+d_q\right)}
=\frac{1}{\Gamma(d_q-c_q)}\int_0^1 dy_q \, y_q^{m+c_q-1}
(1-y_q)^{d_q-c_q-1}
,
\\&&
\frac{\Gamma\left(\ell+m+e'_r \right)}{\Gamma\left(\ell+m+f'_r\right)}
=\frac{1}{\Gamma(f'_r-e'_r)}\int_0^1 dz_r \, z_r^{\ell+m+e'_r-1}
(1-z_r)^{f'_r-e'_r-1}
.
\label{Beta3}
\eea
Hence, we obtain an integral representation
\bea
&&
K\propto \int_0^1\!\!dx_1\cdots\int_0^1\!\!dx_L
~
\int_0^1\!\!dy_1\cdots\int_0^1\!\!dy_M
~
\int_0^1\!\!dz_1\cdots\int_0^1\!\!dz_N
~
\nonumber\\&&
~~~~~~~
\times 
\frac{
\prod_p  x_p^{a_p-1} (1-x_p)^{b_p-a_p-1}
\prod_q  y_q^{c_q-1} (1-y_q)^{d_q-c_q-1}
\prod_r  z_r^{e'_r-1} (1-z_r)^{f'_r-e'_r-1}
}
{(1-x_1\cdots x_L z_1\cdots z_N)(1-y_1\cdots y_M z_1\cdots z_N)}
.
\nonumber\\&&
\label{IntegRepK}
\eea

\noindent
\subsubsection*{Step 2}

\begin{figure}[t]\centering
\includegraphics[width=8cm]{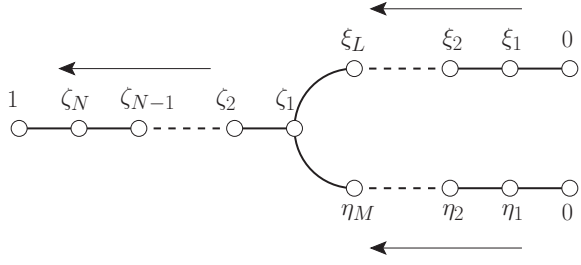}
\caption{\small
\label{FigVarOrder}
Diagram showing the order of the integral variables.
Along the same line, the variables to the left are
larger.
The order of $\xi_p$ and
$\eta_q$ is not fixed.
}
\end{figure}

We apply Kummer's formula
\bea
\int_0^1 \! dt\, \frac{t^\alpha(1-t)^\beta}{(1-tT)^\gamma}
=\frac{\Gamma(\alpha+1)\Gamma(\beta+1)}{\Gamma(\gamma)
\Gamma(\alpha\!+\!\beta\!-\!\gamma\!+\!2)}
\frac{(1-T)^{\beta-\gamma+1}}{T^{\alpha+\beta-\gamma+2}}
\int_0^T\!\! dt \,\frac{t^{\alpha+\beta-\gamma+1}}{(1-t)^{\beta+1}}
\biggl(1\!-\!\frac{t}{T}\biggr)^{\!\gamma-1}
\nonumber\\
\label{Kummer}
\eea
for $\gamma=1$ to the variables $t=x_1$ and $y_1$ of
eq.~(\ref{IntegRepK}).
Next we apply the following variable transformations
successively (here, $\zeta_i$'s denote the new variables and not
zeta values):
\begin{itemize}
\item
$z_N=\zeta_N$, and
$z_i=\zeta_i/\zeta_{i+1}$ starting from $i=N-1$ to $i=1$.\footnote{
If $N=1$, $z_1=\zeta_1$.
}
\item
$x_1=\xi_1$, $x_L=\xi_L/\zeta_1$, and
$x_i=\xi_i/\xi_{i+1}$ starting from $i=L-1$ to $i=2$.\footnote{
If $L=1$, $x_1=\xi_1$;
if $L=2$, $x_1=\xi_1$ and $x_2=\xi_2/\zeta_1$.
Similar transformations apply in the cases $M=1,2$.
}
\item
$y_1=\eta_1$, $y_M=\eta_M/\zeta_1$, and
$y_i=\eta_i/\eta_{i+1}$ starting from $i=M-1$ to $i=2$.
\end{itemize}
This leads to an integral representation of $K$, 
where the integral variables
are ordered as depicted in Fig.~\ref{FigVarOrder}, and where
each factor of the integrand
contains at most two (adjacent) integral variables:
\bea
&&
K  \propto
{\hbox to 18pt{
\hbox to -3pt{$\displaystyle \int$} 
\raise-18pt\hbox{$\scriptstyle 1\geq \zeta_N\geq
\cdots \geq \zeta_1\geq 0$} 
}}
\!d\zeta_{N}\cdots d\zeta_{1}~
{\hbox to 18pt{
\hbox to -3pt{$\displaystyle \int$} 
\raise-18pt\hbox{$\scriptstyle 
\zeta_1\geq \xi_L\geq\cdots\geq \xi_1\geq 0$} 
}}
\!d\xi_{L}\cdots d\xi_{1}~
{\hbox to 18pt{
\hbox to -3pt{$\displaystyle \int$} 
\raise-18pt\hbox{$\scriptstyle 
\zeta_1\geq \eta_M\geq\cdots\geq \eta_1\geq 0$} 
}}
\!d\eta_{M}\cdots d\eta_{1}~~
\zeta_{1}^{e'_{1}-a_{L}-c_{M}-1}
\left(1-\zeta_{N}\right)^{f'_{N}-e'_{N}-1}
\nonumber \\
 && \qquad\times
\left(1\!-\!\xi_{1}\right)^{a_{1}-b_{1}}\xi_{1}^{b_{1}-2}\left(1\!-\!\xi_{2}\right)^{b_{1}-a_{1}-1}\xi_{2}^{a_{2}-b_{1}}
\left(1\!-\!\eta_{1}\right)^{c_{1}-d_{1}}\eta_{1}^{d_{1}-2}\left(1\!-\!\eta_{2}\right)^{d_{1}-c_{1}-1}\eta_{2}^{c_{2}-d_{1}}
\rule[0mm]{0mm}{6mm}
\nonumber \\
 && \qquad\times
\left(1-\frac{\xi_{L}}{\zeta_{1}}\right)^{b_{L}-a_{L}-1}\left(1-\frac{\eta_{M}}{\zeta_{1}}\right)^{d_{M}-c_{M}-1}
\rule[0mm]{0mm}{8mm}
\nonumber \\
 && \qquad\times
\left(1-\frac{\xi_{2}}{\xi_{3}}\right)^{b_{2}-a_{2}-1}\xi_{3}^{a_{3}-a_{2}-1}\cdots\left(1-\frac{\xi_{L-1}}{\xi_{L}}\right)^{b_{L-1}-a_{L-1}-1}\xi_{L}^{a_{L}-a_{L-1}-1}\nonumber \\
 && \qquad\times
\left(1-\frac{\eta_{2}}{\eta_{3}}\right)^{d_{2}-c_{2}-1}\eta_{3}^{c_{3}-c_{2}-1}\cdots\left(1-\frac{\eta_{M-1}}{\eta_{M}}\right)^{d_{M-1}-c_{M-1}-1}\eta_{M}^{c_{M}-c_{M-1}-1}\nonumber \\
 && \qquad\times
\left(1-\frac{\zeta_{1}}{\zeta_{2}}\right)^{f'_{1}-e'_{1}-1}\zeta_{2}^{e'_{2}-e'_{1}-1}\cdots\left(1-\frac{\zeta_{N-1}}{\zeta_{N}}\right)^{f'_{N-1}-e'_{N-1}-1}\zeta_{N}^{e'_{N}-e'_{N-1}-1}
.
\label{LengthyExp:K}
\eea
We note that all the factors containing two variables have
integer powers at $\epsilon=0$, 
according to the condition (ii) and eq.~(\ref{efPrime}).
On the other hand, the factors $t$ and $1-t$ for
$t=\xi_p,\eta_q,\zeta_r$ have half-integer (including integer)
powers at $\epsilon=0$.

We may render the exponents of
all the factors of the integrand to integers
(at $\epsilon=0$) using
the following variable transformations.
We first apply the transformation
\bea
t\rightarrow\frac{4t}{\left(1+t\right)^{2}}.
\label{VarTransfElimHI}
\eea
to all the variables $t=\xi_p,\eta_q,\zeta_r$ simultaneously.
The integral region is unchanged, and the factors
in the integrand transform as
\bea
1-t\rightarrow\left(\frac{1-t}{1+t}\right)^2,\qquad1-\frac{t}{t'}\rightarrow\frac{\left(1-{t}/{t'}\right)\left(1-t\,t'\right)}{t'\left(1+t\right)^{2}} .
\eea
If half-integer powers still remain, we further transform
$t\to t^2$.
This eliminates half-integer powers in the general case.

\noindent
\subsubsection*{Step 3}

Although $K$ is finite at $\epsilon = 0$, 
setting $\epsilon= 0$ in the integrand before
integration may lead to divergences.
These superficial divergences originate from
the integral representations eqs.~(\ref{Beta1})--(\ref{Beta3}) and
(\ref{Kummer}), 
in the case that some of
these integrals are not well-defined around $\epsilon=0$.
If there are 
(superficial) divergences, we introduce 
regularization parameters in the following manner,
before expanding in $\epsilon$:
\bea
&&
a_{i}  \rightarrow a_{i}+\delta_{a_{1}}+\cdots+\delta_{a_{i}} , 
~~
\qquad b_{i}\rightarrow b_{i}+\delta_{a_{1}}+\cdots+\delta_{a_{i}}+\delta_{b_{i}} ,
\nonumber
\\&&
c_{i}  \rightarrow c_{i}+\delta_{c_{1}}+\cdots+\delta_{c_{i}} ,
~~~
\qquad d_{i}\rightarrow d_{i}+\delta_{c_{1}}+\cdots+\delta_{c_{i}}+\delta_{d_{i}} ,
\nonumber
\\&&
e'_{i}  \rightarrow e'_{i}+\delta_{a_{1}}+\cdots+\delta_{a_{L}}+\delta_{c_{1}} 
+\cdots+\delta_{c_M}+\delta_{e'_{1}}+\cdots+\delta_{e'_{i}} ,
\nonumber
\\&&
f'_{i}  \rightarrow f'_{i}+\delta_{a_{1}}+\cdots+\delta_{a_{L}}+\delta_{c_{1}} 
+\cdots+\delta_{c_M}+\delta_{e'_{1}}+\cdots+\delta_{e'_{i}}+\delta_{f'_{i}} .
\label{RegParams}
\eea
This parametrization is specially tailored for the use in step 4.
As we will see,
it is important that we can take
the limit $\delta_i\to0$ corresponding to each integral variable
(each set of integral variables),
without affecting outer integrals.
In fact, after applying the above regularization,
most exponents of the factors of the integrand
depend on only one regularization parameter, respectively.
For reference, in Table~\ref{tabPowers} we list  
the exponents in eq.~(\ref{LengthyExp:K}) after applying
the regularization,
from which one may easily deduce the integral form
after the variable transformation eq.~(\ref{VarTransfElimHI})
(and $t\to t^2$).
\begin{table}[t]
\centering
\begin{tabular}{llll}
\hline
\footnotesize
$ {e'_{1}-a_{L}-c_{M}-1+\delta_{e'_{1}}},
$&\footnotesize
${f'_{N}-e'_{N}-1+\delta_{f_{N}}},
~~~~~~
$&\footnotesize
${a_{1}-b_{1}-\delta_{b_{1}}},
$&\footnotesize
${b_{1}-2+\delta_{a_{1}}+\delta_{b_{1}}},
$
\\
\footnotesize
$ {b_{1}-a_{1}-1+\delta_{b_{1}}},
$&\footnotesize
${a_{2}-b_{1}+\delta_{a_{2}}-\delta_{b_{1}}},
$&\footnotesize
$ {c_{1}-d_{1}-\delta_{d_{1}}},
$&\footnotesize
${d_{1}-2+\delta_{c_{1}}+\delta_{d_{1}}},
$
\\
\footnotesize
$  {d_{1}-c_{1}-1+\delta_{d_{1}}},
$&\footnotesize
$ {c_{2}-d_{1}+\delta_{c_{2}}-\delta_{d_{1}}},
$&
\footnotesize
$ {b_{L}-a_{L}-1+\delta_{b_{L}}},
$
&\footnotesize
$ {d_{M}-c_{M}-1+\delta_{d_{M}}},
$
\\
\footnotesize
$ {b_{2}-a_{2}-1+\delta_{b_{2}}},
$&\multicolumn{2}{l}{
\footnotesize
$ {a_{3}-a_{2}-1+\delta_{a_{3}}},
~\cdots,~ {b_{L-1}-a_{L-1}-1+\delta_{b_{L-1}}},
$
}
&\footnotesize
$ {a_{L}-a_{L-1}-1+\delta_{a_{L}}},
$
\\
\footnotesize
$ {d_{2}-c_{2}-1+\delta_{d_{2}}},
$&\multicolumn{2}{l}{
\footnotesize
$ {c_{3}-c_{2}-1+\delta_{c_{3}}},
~\cdots,~
{d_{M-1}-c_{M-1}-1+\delta_{d_{M-1}}},
$
}
&\footnotesize
$ {c_{M}-c_{M-1}-1+\delta_{c_{M}}},
$
\\
\footnotesize
$ {f'_{1}-e'_{1}-1+\delta_{f'_{1}}},
$&\multicolumn{2}{l}{
\footnotesize
$ {e'_{2}-e'_{1}-1+\delta_{e'_{2}}},
~\cdots,~
{f'_{N-1}-e'_{N-1}-1+\delta_{f'_{N-1}}},
$
}
&\footnotesize
$ {e'_{N}-e'_{N-1}-1+\delta_{e'_{N}}}.
$
\\
\hline
\end{tabular}
\caption{\small
\label{tabPowers}
Exponents of the factors of the integrand in eq.~(\ref{LengthyExp:K})
after introducing regularization parameters
eq.~(\ref{RegParams}).
The exponents of $\xi_1,\xi_2,\eta_1,\eta_2$, respectively, depend on
two regularization parameters.
All other
exponents,
respectively, depend on a single regularization parameter.
}
\end{table}

We expand
$K$ in $\epsilon$ 
(if necessary,
after applying the above regularization).
The expansion generates powers of
logarithms 
[$\log t$, $\log(1\pm t)$, $\log(1\pm it)$, $\log(1\pm t/t')$,
$\log(1\pm t\,t')$] in the integrand.

\noindent
\subsubsection*{Step 4}

We convert the integral to a sum of nested 
integrals using a recursive algorithm.
The order of conversion is indicated by the arrows in 
Fig.~\ref{FigVarOrder}: we work
from the innermost integral up to the outermost integral.
First we convert the double integrals
with respect to the innermost variables
$\xi_1$,$\xi_2$ and $\eta_1$,$\eta_2$, respectively.
For the other variables,
we convert the integral
with respect to each variable recursively.\footnote{
The cases with $L=1$ or $M=1$ are exceptional.
If $L=1$ and $M\geq 2$, we work in the following order:
we first convert the double integral with respect to
$\eta_1,\eta_2$;
we work along the $\eta$ chain for each variable
recursively up to $\eta_M$;
we convert the double integral with respect to 
$\xi_1$ and $\zeta_1$;
we work along the $\zeta$ chain recursively up to $\zeta_N$.
(If $M=1$ and $L\geq 2$, exchange $\xi_i$ and $\eta_i$.)
If $L=M=1$ we convert the triple integral
with respect to $\xi_1,\eta_1,\zeta_1$ first;
the method is similar to the double integral case.
}

Let us explain how to convert the double integral with respect to
$\xi_1$ and $\xi_2$.
The integrand depends on the  four regularization parameters
$\delta_{a_{1}}$,$\delta_{a_{2}}$,$\delta_{b_{1}}$,$\delta_{b_{2}}$
(denoted by $\delta_{\xi 12}$).
We note that these parameters are not included in the outer
integrals. 
Using identities and integration by parts,
we raise or decrease the exponents of the factors,
$\xi_i, 1\!\pm\!\xi_i, 1\!\pm\!i\xi_i$ ($i=1,2$),
$1\!\pm\!\xi_2/\xi_3, 1\!\pm\!\xi_2 \xi_3
$,
until all of them become non-negative 
when $\delta_{\xi 12}$'s are set to zero.
This procedure is explained in App.~\ref{appB}.
By this procedure,
the double integral is converted to a
sum of double integrals, where
singularities in $\delta_{\xi 12}$'s are taken outside of
each double integral
as poles and each double integral is finite as all $\delta_{\xi 12}\to 0$.
We expand $K$ in $\delta_{\xi 12}$'s up to (and including)
finite terms.
The poles in $\delta_{\xi 12}$'s cancel, and each 
double integral with respect
to $\xi_1$ and $\xi_2$ is reduced to the form
\bea
&&
\int_0^{\xi_3}\!\!d\xi_2\int_0^{\xi_2}\!\!d\xi_1~ 
P_{\xi 12}\Bigl(\xi_i,
\log\xi_i, \log(1\!\pm\!\xi_i), \log(1\!\pm\!i\xi_i),
 \log(1\!\pm\!\xi_2/\xi_3), \log(1\!\pm\!\xi_2 \xi_3)
\Bigr)
\nonumber\\
\eea
where $P_{\xi 12}$ is a polynomial of its arguments.
Prefactors dependent on $\xi_3$ may be generated,
which are absorbed into the outer integral $\int d\xi_3$.
The above integral can be converted to a
combination of HPLs
using Algorithm I, as described in Sec.~\ref{s3a}.\footnote{
All the logs in $P_{\xi 12}$ can be expressed
by an HPL.
It is useful to use the shuffle relations 
(fusion rule) of HPLs to express their
products as a sum of HPLs.
}
In fact, it is a generalized version of
eq.~(\ref{ExI1}).

In the same way, we convert the double integral
with respect to $\eta_1$ and $\eta_2$ to a sum of HPLs.

Next we apply the following algorithm recursively
from inner to outer integrals.
Suppose we convert
the integral 
with respect to $t(=\xi_{p(\geq 3)},\eta_{q(\geq 3)},\zeta_r$).
As a result of the previous conversion,
each integral has the following form:
\bea
&&
\int_0^{t'}\!dt~ 
\left[\prod_{s=1}^{9} \phi_s(t,t')^{\alpha_s(\delta_1,\delta_2)}
\right]
~
{P}\Bigl(
\log\phi_1(t,t'),\dots,\log\phi_{9}(t,t') 
\Bigr)
\, H(t).
\label{originalform}
\eea
Here, $\phi_s(t,t')$ denotes one of the nine factors
$ t$, $(1\pm t)$, $(1\pm it)$, $(1\pm t/t')$,
$(1\pm t\,t')$.
$P(x_1,\dots,x_9)$ represents a polynomial of $x_1,\dots,x_9$. 
$H(t)$ is
a nested integral which resulted from processing
the inner integrals.\footnote{
For $t=\zeta_1$, $H(t)$ is replaced by 
a product of $H_\xi(t)$ and $H_\eta(t)$, which
originate from $\xi$ and $\eta$ chains, respectively;
see Fig.~\ref{FigVarOrder}.
One can use the shuffle relations 
(fusion rule) to express $H_\xi(t) H_\eta(t)$ as a sum of 
nested integrals.
}
In fact, 
$H(t)$ is a generalized HPL of eqs.~(\ref{HPL1})--(\ref{fp}).
There are only
two regularization parameters
associated with the variable $t$, and we denote them by
$\delta_1,\delta_2$.
In particular, $\delta_1,\delta_2$ are not included in the outer
integrals. 
The exponent
$\alpha_s(\delta_1,\delta_2)$ is a linear function of
$\delta_1,\delta_2$, where $\alpha_s(0,0)$ is an integer.

We raise or decrease
the exponents $\alpha_s(\delta_1,\delta_2)$'s 
using identities and integration by parts (see App.~\ref{appB}).
By this procedure,
the original integral is converted to a
combination of integrals, where
singularities in $\delta_1,\delta_2$ are taken outside of each integral
as poles and each integral is finite as $\delta_1,\delta_2\to 0$.
We expand $K$ in $\delta_1,\delta_2$ up to (and including)
finite terms.
The poles in $\delta_1,\delta_2$ cancel, and each integral with respect
to $t$ is reduced to the form
\bea
&&
\int_0^{t'}\!dt~ 
\tilde{P}\Bigl(
t,\, \log\phi_1(t,t'),\,\dots,\,\log\phi_{9}(t,t') 
\Bigr)
\, H(t),
\eea
where $\tilde{P}$ is a polynomial of its arguments.
Prefactors dependent on $t'$ may be generated,
which should be absorbed into the outer integral.\footnote{
We note that $c_1(t')$ and $c_2(t')$
in eq.~(\ref{decompone}) can be expressed by
products of powers of $t'$, $(1\pm t')$, $(1\pm it')$.
For this reason, the prefactors 
can also be expressed by these factors.
Thus, the absorption of the prefactor does not alter the
form eq.~(\ref{originalform}).
}
The above integral can be converted to a
combination of nested integrals
using Algorithm I, as described in Sec.~\ref{s3a}.

Finally, since the upper bound of the outermost integral is
one, the result can be expressed by MZVs with
roots of unity ($\beta_i^{P_i}=1$ with $P_i\in\mathbb{N}$).

\noindent
\subsubsection*{Step 5}

The same as step 4 of Algorithm I (Sec.~\ref{s2}).

\section{Applications of Algorithm II}
\label{s5}
\clfn

\begin{figure}[t]\centering
\includegraphics[width=5cm]{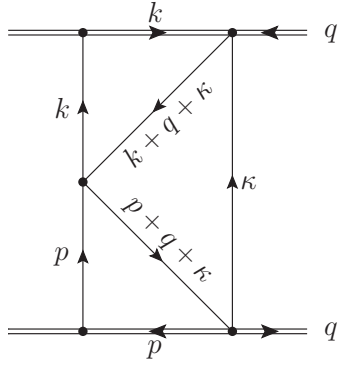}
\caption{\small
Diagram for the master integral ${\cal M}_2$.
Notations are the same as in Fig.~\ref{FigM33}.
\label{FigM25}
}
\end{figure}

We apply Algorithm II for evaluating a 3-loop
integral 
\bea
{\cal M}_2=\int \frac{d^Dp}{(2\pi)^D}
\frac{d^Dk}{(2\pi)^D}\frac{d^D\kappa}{(2\pi)^D}\,
\frac{1}{(k\cdot v)(p\cdot v)k^2p^2\kappa^2
(k+q+\kappa)^2(p+q+\kappa)^2} .
\label{M25}
\eea
This is another master integral necessary for a computation of
$\bar{a}_3$;
the diagram is shown in Fig.~\ref{FigM25}.
The notations are the same as in Sec.~\ref{s3c}.

Using Mellin-Barnes 
integral representation, one can express
${\cal M}_2$ by $K$ 
with five different 
arguments with $(L,M,N)=(1,1,2)$.
After expanding in $\epsilon$, six $\Gamma$ functions
remain uncanceled in the summand, for each $K$.
We need to compute
the expansion coefficients of $K$ around three different arguments
$(
a_{0,1},b_{0,1};c_{0,1},d_{0,1};e_{0,1},e_{0,2},f_{0,1},f_{0,2}
)
=(\frac{1}{2},1;\frac{1}{2},1;\frac{1}{2},1,1,2)
$,
$(\frac{1}{2},1;1,\frac{3}{2};1,\frac{3}{2},\frac{3}{2},\frac{5}{2})
$, 
$(1,\frac{3}{2};1,\frac{3}{2};\frac{3}{2},2,2,3)
$ up to order $\epsilon^5$, $\epsilon^3$, $\epsilon^3$, 
respectively.
We can use Algorithm II to evaluate these coefficients
and obtain
\bea
&&
{\cal M}_2=A(\epsilon)(-q^2)^{-3\epsilon}\times
\Biggl[
-\frac{28 \pi ^4}{135 \epsilon }
+\left\{
-\frac{226 \zeta (5)}{3}-\frac{116 \pi ^2 \zeta (3)}{9}+\pi ^4 \left(4 \log 2-\frac{224}{135}\right)
\right\}
\nonumber\\&&~~~~~
+\epsilon ~ \biggl\{192 {s_6}-\frac{1808 \zeta (5)}{3}+\frac{8 \zeta (3)^2}{3}
+\pi ^2 \left(-64 \text{Li}_4\bigl({\textstyle \frac{1}{2}}\bigr)-\frac{928 \zeta (3)}{9}-\frac{8}{3} \log ^4 2\right)
\nonumber\\&&~~~~~
~~~~~~~~
+\pi ^4 \left(-\frac{1792}{135}+\frac{20 \log^2 2}{3}+32 \log 2\right)
+\frac{428 \pi ^6}{2835}
\biggr\}
\nonumber\\&&~~~~~
+
\epsilon ^2 \biggl\{768 \text{Li}_4\bigl({\textstyle \frac{1}{2}}\bigr) \zeta (3)
+\frac{480}{7} \zeta (3)^2 \log 2
+1536 {s_6}-\frac{384}{7} {s_6} \log 2+\frac{384}{7} s_{7a}
+\frac{3072}{7} s_{7b}
\nonumber\\&&~~~~~
~~~~~~~~
-\frac{4960 \zeta (7)}{21}-\frac{14464 \zeta (5)}{3}+\frac{64 \zeta (3)^2}{3}+32 \zeta (3) \log ^4 2-372 \zeta (5) \log^2 2
\nonumber\\&&~~~~~
~~~~~~~~
+\pi ^2 \biggl(-512 \text{Li}_4\bigl({\textstyle \frac{1}{2}}\bigr)+128 \text{Li}_5\bigl({\textstyle \frac{1}{2}}\bigr)-\frac{7424 \zeta (3)}{9}
-\frac{35519 \zeta (5)}{42}-32 \zeta (3) \log^2 2
\nonumber\\&&~~~~~
~~~~~~~~
-\frac{16 \log ^5 2}{15}-\frac{64}{3} \log ^4 2\biggr)
+\pi ^4 \biggl(\frac{31457 \zeta (3)}{945}-\frac{14336}{135}+\frac{40 \log ^3 2}{9}
\nonumber\\&&~~~~~
~~~~~~~~
+\frac{160 \log^2 2}{3}+256 \log 2\biggr)
+\pi ^6 \left(\frac{3424}{2835}+\frac{3079 \log 2}{315}\right)\biggr\}
\Biggr] ,
\eea
where 
$s_{7a}=z(-5,1,1)+z(-6,1)+z(-5,2)+z(-7)
=\sum_{m\geq n\geq k\geq 1}\frac{(-1)^m}{m^5nk}
=-0.9529600757562\dots
$ and
$s_{7b}=z(5,-1,-1)+z(7)+z(5,2)+z(-6,-1)
=\sum_{m\geq n\geq k\geq 1}\frac{(-1)^{n+k}}{m^5nk}
=1.029121262964324\dots
$.
We have checked the above result 
numerically by evaluating Feynman parameter integral
representation using sector decomposition.
This integral corresponds to $I_{14}$ given in 
\cite{Smirnov:2010zc},\footnote{
Ref.~\cite{Smirnov:2010zc} presents analytical results for
the expansion coefficients of (selected) 16 
master integrals for $\bar{a}_3$.
} 
where ${\cal O}(\epsilon^2)$ term is missing.\footnote{
While preparing for this paper, the ${\cal O}(\epsilon^2)$
term has been computed 
in \cite{Lee:2012te} using 
the dimensional recurrence and analyticity 
method in combination with \texttt{PSLQ}
algorithm \cite{pslq}
(a sophisticated estimate of analytical results
from high-precision numerical results).
Their result and our result are in agreement.
In comparison to their method, our method of computation
is fully analytic.
}

We have computed yet
another master integral for $\bar{a}_3$ using Algorithm II.
This is $I_{16}$ given in \cite{Smirnov:2010zc},
where the analytical
expressions for the
coefficients of the $\epsilon^{-1}$ and $\epsilon^{0}$
terms are presented.
Ref.~\cite{Smirnov:2010zc} uses
the Mellin-Barnes method in combination with \texttt{PSLQ}
algorithm for the computation.
We have reproduced their result, thus providing a
cross check in a fully analytical way.

\section{Summary and Discussion}
\label{s6}

We have constructed algorithms for computing
two types of multiple sums (Algorithms I and II)
which appear in higher-order
loop computations.
For instance, these types of sums appear in 
analytic evaluations
of Mellin-Barnes integral representations by
closing contours and expanding in $\epsilon$.

Algorithm I applies to a multiple sum
without $\Gamma$ functions [eq.~(\ref{SumAlg1})]:
\bea
S(\nu) = 
\sum_{i_1=a_1(\nu)}^{b_1(\nu)}
\sum_{i_2=a_2(\nu,i_1)}^{b_2(\nu,i_1)}
\cdots \sum_{i_N=a_N(\nu,i_1,\dots,i_{N-1})}^
{b_N(\nu,i_1,\dots,i_{N-1})}
~
\frac{\lambda_1^{i_1}\cdots \lambda_N^{i_N}}
{\prod_r L_r(\nu,i_1,\dots ,i_N)^{p_r}} .
\label{}
\eea
The algorithm is valid for arbitrary multiplicity $N$
and works efficiently.
(This is partly due to our specific regularization
scheme for treating surface terms at infinity;
see discussion at the end of Sec.~2.)
The results are expressed by $Z$-sums 
[with argument $n\nu$ ($n\in\mathbb{N}$)] and 
generalized MZVs.

There are benefits for constructing an efficient algorithm
specifically for this type of sums.
First, this type of sums appear frequently in 
computation of expansion coefficients of
loop integrals in $\epsilon$.
Secondly, it has other useful applications as shown
in Sec.~\ref{s3}:
(i)~Conversion of non-nested integrals to nested integrals
(used in Algorithm II),
(ii)~Deriving non-trivial relations among MZVs;
for example, we find a further reduction of the
proposed basis elements at $(w,l)=(2,8)$ in \cite{Ablinger:2011te}.

Algorithm II is used to evaluate the expansion coefficients
of a double sum with $\Gamma$ functions
[eq.~(\ref{eq:K})],
\bea
K(\boldsymbol{a},\!\boldsymbol{b};\boldsymbol{c},\!\boldsymbol{d};
\boldsymbol{e},\!\boldsymbol{f})
=
\sum_{\ell,m=0}^{\infty}
\Biggl[\prod_{p=1}^L
\frac{\Gamma\left(\ell\! + \!a_p\right)}{\Gamma\left(\ell\! + \!b_p\right)}
\Biggr]
\Biggl[\prod_{q=1}^M
\frac{\Gamma\left(m\! + \!c_q\right)}{\Gamma\left(m\! + \!d_q\right)}
\Biggr]
\Biggl[\prod_{r=1}^N
\frac{\Gamma\left(\ell\! + \!m\! + \!e_r \right)}{\Gamma\left(\ell\! + \!m\! + \!f_r\right)}
\Biggr]
,
\label{eq:K2}
\eea
around half-integer values 
(including integer values) of its arguments
$(\boldsymbol{a}_0,\!\boldsymbol{b}_0;\boldsymbol{c}_0,\!\boldsymbol{d}_0;
\boldsymbol{e}_0,\!\boldsymbol{f}_0)$.
For $p,q,r\geq 2$, $a_{0,p}-b_{0,p}$, $c_{0,q}-d_{0,q}$, 
$e_{0,r}-f_{0,r}$ are assumed to be integers.
Hence, there remain at most six uncanceled
$\Gamma$ functions in the summand in
each expansion coefficient.
The algorithm applies to arbitrary expansion coefficients.
The results are expressed by generalized MZVs
with roots of unity.

These algorithms are useful in evaluating some complicated
master integrals which appear in computation
of the 3-loop correction to the
static QCD potential ($\bar{a}_3$).
We have presented new results for two master integrals
(Secs.~\ref{s3c} and \ref{s5}).

In practical implementation of the algorithms\footnote{
A {Mathematica} package ${\it ``Wa"}$
for computing multiple sums using the
algorithms developed in this paper is available
at 
\texttt{http://www.tuhep.phys.tohoku.ac.jp/$\sim$program/} 
with examples and instructions.
In particular, Type I sums can be evaluated fairly efficiently.
},
one can make various improvements to compute efficiently.
For Algorithm I, however, at the moment we do not have ideas for
essential modifications of the algorithm, apart from
technical refinements.
On the other hand, for Algorithm II,
we may improve efficiencies non-trivially
by categorizing the arguments of $K$ and
changing algorithms
according to the categories.
Although in most difficult cases $K$
can be evaluated only by
the algorithm described in Sec.~\ref{s4},
in simpler cases, it is possible to construct algorithms which
work substantially faster:
for instance, we may express $K$ in an integral representation
involving hypergeometric functions $_2F_1$ and utilize
various identities of $_2F_1$.
We explain details of our practical implementation elsewhere.

In computation of $\bar{a}_3$
we encounter multiple sums which cannot be evaluated
with Algorithms I and II.
An example is a sum, where
the arguments of $\Gamma$ functions in the numerator
and denominator include the summation indices
in different combinations, such as:
\bea
\sum_{k,m,n,\cdots}
\frac{\Gamma (m+a) \Gamma (n+b) \Gamma (k+m+c) \Gamma
 (k+m+n+d)\cdots}
{\Gamma (k+a') \Gamma (m+n+b') \Gamma (k+2 m+n+c')\cdots}
.
\eea
To our knowledge, there are no known technologies which
can be used to evaluate these sums in general cases.
We are currently studying how to
generalize the algorithms 
in order to evaluate these
different types of sums.
It may be useful to
combine the present algorithms
with other methods for analytical evaluation of
loop integrals,
such as the method of differential equation, glue-and-cut method, 
etc.


\appendix
\clfn
\section*{Appendices}

\section{Definitions and Conventions}
\label{appA}

$Z$-sums are defined by
\bea
&&
Z\left(w;\alpha_{1},\alpha_{2},\ldots,\alpha_{n};\beta_{1},\beta_{2},\ldots,\beta_{n}\right) =\sum_{w\geq k_{1}>k_{2}>\cdots>k_{n}\geq1}\frac{\beta_{1}^{k_{1}}\beta_{2}^{k_{2}}\cdots\beta_{n}^{k_{n}}}{k_{1}^{\alpha_{1}}k_{2}^{\alpha_{2}}\cdots k_{n}^{\alpha_{n}}}
\nonumber\\&&
~~~~~~~~~~~~~~~~~~~~~~~~~~~~~~~~~~~~~~~~~~~~
=\sum_{k_{1}=1}^{w}\frac{\beta_{1}^{k_{1}}}{k_{1}^{\alpha_{1}}}\sum_{k_{2}=1}^{k_{1}-1}\frac{\beta_{2}^{k_{2}}}{k_{2}^{\alpha_{2}}}\cdots\sum_{k_{n}=1}^{k_{n-1}-1}\frac{\beta_{n}^{k_{n}}}{k_{n}^{\alpha_{n}}},
\eea
where $\alpha_i, w \in\mathbb{N}$ and $\beta_i\in \mathbb{C}$.
For $w=\infty$, the sums are (generalized) multiple zeta values\footnote{
It is more common to restrict to the case $\beta_i=\pm 1$
when referring to MZVs.
In view of applications to higher-order loop computations,
where this restriction becomes inconvenient,
we relax conditions on $\beta_i$ in this paper.
}
(MZVs) or multiple polylogarithms of Goncharov:
\bea
Z\left(\infty;\alpha_{1},\alpha_{2},\ldots,\alpha_{n};\beta_{1},\beta_{2},\ldots,\beta_{n}\right)=\mathrm{Li}_{\alpha_{n},\ldots,\alpha_{1}}\left(\beta_{n},\ldots,\beta_{1}\right) .
\label{MZV}
\eea
We also use a short-hand notation for MZVs, 
in contexts where it is possible to
restore the original forms:
\bea
Z(\infty;\alpha_1,\dots,\alpha_n;\beta_1,\dots,\beta_n)
=z(\alpha_1\beta_1,\dots,\alpha_n\beta_n).
\label{short-hand-not}
\eea

Harmonic polylogarithms (HPLs) are defined 
recursively:
For weight one, they are defined by\footnote{
The definition in this paper differs from the conventional
one, $f_{\pm 1}(x)=\frac{1}{1\mp x}$,
considering the generalization eq.~(\ref{fp}).
}
\bea
&&
f_{0}\left(x\right)=\frac{1}{x},\qquad 
f_{\pm1}\left(x\right)=\frac{1}{x\mp1},
\label{HPL1}
\\ &&
H_0\left(x\right) =\log\left(x\right),\qquad 
H_{\pm1}\left(x\right) =\int_{0}^{x}\!dt \,\,
f_{\pm1}\left(t\right)=\log\left(1\mp x\right),
\eea
and for higher weights,
\bea
&&
H_{\underbrace{\scriptstyle 0,\cdots,0}_{n}}(x)  =\frac{1}{n!}\left[\log\left(x\right)\right]^{n} ,
~~~~
H_{a_{1},a_{2},\ldots,a_{n}}
(x) =\int_{0}^{x}\!dt \,\,
f_{a_{1}}(t)H(t;a_{2},\ldots,a_{n}) ,
\label{HPL}
\eea
where at least one of $a_{1},a_{2},\ldots,a_{n}$ is non-zero.
The above definition can be generalized  to the case
\bea
f_{p}(x)=\frac{1}{x-p}
~~~;~~~
p \in \mathbb{C}
\label{fp}
\eea
where $a_i$ of $H_{a_{1},a_{2},\ldots,a_{n}}(x)$ becomes
a general complex number.
We refer to it as a generalized HPL.
$H_{a_{1},a_{2},\ldots,a_{n}}(x)$ is another representation of an
MZV or multiple polylogarithm
[eq.~(\ref{MZV})].\footnote{
The summation representation is obtained 
by first writing the integrand of nested
integral representation in Taylor series and then integrating.
}

\section{Procedure for Raising or Decreasing Exponents}
\label{appB}

In this appendix we explain the procedure used
in step 4 of Algorithm II (see Sec.~\ref{s4}).
It is used to raise or
decrease the exponents of the factors in the integrand,
until all of them become non-negative
when the regularization parameters are set to zero.

We consider an integral of the form
\bea
&&
\int_0^{t'}\!dt~ 
\left[\prod_{s=1}^{9} \phi_s(t,t')^{\alpha_s(\boldsymbol{\delta})}
\right]
~
{P}\Bigl(
\log\phi_1(t,t'),\dots,\log\phi_{9}(t,t') 
\Bigr)
\, H(t).
\label{AppB-int1}
\eea
Here, $\phi_s(t,t')$ denotes one of the nine factors
$ t$, $(1\pm t)$, $(1\pm it)$, $(1\pm t/t')$,
$(1\pm t\,t')$;
$P(x_1,\dots,x_9)$ represents a polynomial of $x_1,\dots,x_9$;
$H(t)$ is a generalized HPL of eqs.~(\ref{HPL1})--(\ref{fp});
$\boldsymbol{\delta}=(\delta_1,\dots,\delta_n)$ denote
the regularization parameters;
the exponent
$\alpha_s(\boldsymbol{\delta})$ is a linear function of
$\boldsymbol{\delta}$, where $\alpha_s(\boldsymbol{0})$ is an integer.

A rough sketch of the procedure is as follows.
We first multiply the integrand by one,
which can be decomposed as
\bea
&&
1=t+(1-t)=\frac{t}{t'}+\left(1-\frac{t}{t'}\right)
=\frac{1}{1+t'}\,(1-t)+\frac{t'}{1+t'}\left(1+\frac{t}{t'}\right)
,
\eea
etc.
This is repeated until the exponents are raised sufficiently.
Then we integrate by parts to trade 
between different exponents.

Let us explain in detail.
First we select a pair of the factors $\phi_{s}$ and 
$\phi_{\sigma}$, for which $\alpha_s(\boldsymbol{0})$
and $\alpha_\sigma(\boldsymbol{0})$ are both negative.
We multiply the integrand of eq.~(\ref{AppB-int1}) by
\bea
1=c_s(t')\phi_{s}(t,t')+c_\sigma(t')\phi_{\sigma}(t,t').
\label{decompone}
\eea
This is repeated $|\alpha_s(\boldsymbol{0})+
\alpha_\sigma(\boldsymbol{0})|$ times.
After expanding the integrand, in each term, 
the exponents of $\phi_{s}$ and $\phi_{\sigma}$ are
both non-negative or at least one of them is non-negative.\footnote{
It is understood that 
we set $\boldsymbol{\delta}\to\boldsymbol{0}$
when we refer to the signs of the exponents.
}
We repeat this process successively for pairs of negative
exponents in each term.
In the end, in each term, at most one negative exponent
remains.
At the same time, the sum of all the exponents are
non-negative for each term.
Next we perform integration by parts
in every term which has 
$\phi_s^\nu$ with $\nu<0$, in such
a way to integrate 
$\phi_s^\nu$.\footnote{
We note that the derivative of $H(t)$ can be expressed by
a generalized HPL of a lower weight.}\footnote{
Whenever there is a subtlety in 
defining the integrals and surface terms,
they are defined by analytic continuation 
using $\boldsymbol{\delta}$.
}
This is repeated until the exponent of $\phi_s$
becomes non-negative.
In the end, all the exponents can be made
non-negative.

In the case of a double integral with respect to
$\xi_1,\xi_2$,
we iterate the above procedure twice,
first with respect to the inner integral and then the outer integral.
Due to the absence of
the factors $(1\pm\xi_1/\xi_2)$, $(1\pm\xi_1\xi_2)$, iteration of
integration by parts with respect to $\xi_2$ 
can render all the exponents non-negative.
(The same is true for $\eta_1$,$\eta_2$.)

\section*{Acknowledgements}
We are grateful to Y.~Kiyo for fruitful
discussion.
We also thank J.~Bl\"umlein and
C.~Schneider, and 
V.~A.~Smirnov and M.~Steinhauser, respectively, for 
providing information
on their computations.
The works of C.A.\ and Y.S.\ are supported in part by 
JSPS Fellowships for Young Scientists and
Grant-in-Aid for
scientific research (No.\ 23540281) from
MEXT, Japan, respectively.

\end{document}